\documentclass[journal,10pt]{IEEEtran}
\usepackage{ifpdf}
\usepackage{cite}
\usepackage{amsmath}
\usepackage{array}
\usepackage{booktabs}
\usepackage{stfloats}
\usepackage{cite}
\usepackage{graphicx}
\usepackage{amsmath}
\usepackage{epsfig}
\usepackage{epstopdf}
\usepackage{algorithm}
\usepackage{algorithmicx}
\usepackage{algpseudocode}
\usepackage{blkarray}
\usepackage[table,xcdraw]{xcolor}
\usepackage{float}
\usepackage{amsfonts}
\usepackage{subcaption}
\usepackage{mathrsfs}
\usepackage{amssymb}
\usepackage{graphicx}
\usepackage[utf8x]{inputenc}
\usepackage{multirow,bigdelim}
\usepackage{dsfont}
\usepackage{ragged2e}
\usepackage{nicematrix}
\usepackage{mathtools}
\usepackage{hhline}
\usepackage{balance}
\graphicspath{{figures/}}

\DeclareUnicodeCharacter{2212}{\ensuremath{-}}

\begin{document}
	
\title{Joint User and Data Detection in Grant-Free NOMA with Attention-based BiLSTM Network}
	
\author{Saud Khan,~\IEEEmembership{Student Member,~IEEE}, Chandra Thapa,~\IEEEmembership{Member,~IEEE}, Salman Durrani,~\IEEEmembership{Senior Member,~IEEE}, and Seyit Camtepe,~\IEEEmembership{Senior Member,~IEEE
\thanks{Saud Khan and Salman Durrani are with the School of Engineering, The Australian National University, Canberra, 2601, Australia (Email: \{saud.khan, salman.durrani\}@anu.edu.au)
}
\thanks{Saud Khan, Chandra Thapa and Seyit Camtepe are with Data61, Commonwealth Scientific and Industrial Research Organization (CSIRO), Sydney, 2122, Australia (Email: \{chandra.thapa, seyit.camtepe\}@data61.csiro.au)
}
\thanks{This research was undertaken with the assistance of resources and services from the National Computational Infrastructure (NCI), which is supported by the Australian Government. The work from Muhammad Basit Shahab and Sarah J. Johnson was supported by the Australian Government through the Australian Research Council Discovery Projects scheme (Projects DP180100606 and DP210102239).
}
}}


{}%
	
\maketitle

\IEEEpeerreviewmaketitle

\begin{abstract}
We consider the multi-user detection (MUD) problem in uplink grant-free non-orthogonal multiple access (NOMA), where the access point has to identify the total number and correct identity of the active Internet of Things (IoT) devices and decode their transmitted data. We assume that IoT devices use complex spreading sequences and transmit information in a random-access manner following the burst-sparsity model, where some IoT devices transmit their data in multiple adjacent time slots with a high probability, while others transmit only once during a frame. Exploiting the temporal correlation, we propose an attention-based bidirectional long short-term memory (BiLSTM) network to solve the MUD problem. The BiLSTM network creates a pattern of the device activation history using forward and reverse pass LSTMs, whereas the attention mechanism provides essential context to the device activation points. By doing so, a hierarchical pathway is followed for detecting active devices in a grant-free scenario. Then, by utilising the complex spreading sequences, blind data detection for the estimated active devices is performed. The proposed framework does not require prior knowledge of device sparsity levels and channels for performing MUD. The results show that the proposed network achieves better performance compared to existing benchmark schemes.
\end{abstract}

\begin{IEEEkeywords}
Non-orthogonal multiple access, grant-free, deep learning, multi-user detection, internet of things.
\end{IEEEkeywords}

\maketitle

\section{INTRODUCTION}
Grant-free non-orthogonal multiple access (NOMA) is a promising solution to support machine-type communications in the 6G Internet of Things (IoT)~\cite{ letaief2019roadmap}. In traditional grant-based orthogonal multiple access (OMA) schemes, the maximum number of devices being serviced is limited by the number of available orthogonal resources. Therefore, scheduling is required to allow the devices to share the orthogonal resources. In contrast, grant-free NOMA allows devices to transmit their data in an arrive-and-go manner by randomly choosing a resource block without going through the grant-access process \cite{shahab2020grant, shahab2022enabling}. When multiple devices choose the same resource block, a collision occurs, which requires retransmission. These collisions are significantly reduced due to the different multiple access signatures in NOMA \cite{mohammadkarimi2018signature}. Therefore, from a practical perspective, grant-free NOMA is considered an attractive solution for sporadic IoT traffic use cases.

The basic principle of grant-free NOMA is to allow the devices to randomly access the resource blocks through multiple access signatures, such as power levels, spreading sequences, scrambling, and interleaving \cite[Table.~III]{shahab2020grant}. Among these signatures, spreading sequences are considered superior because they can efficiently mitigate multi-user interference \cite{shan2017uplink}. The spreading sequences allow device-specific, low cross-correlation codes to enable grant-free communication. However, in spreading-based signatures, longer-length sequences are needed as the number of devices increases. In this regard, complex spreading sequences, as proposed in multi-user shared access (MUSA) \cite{yuan2016multi}, enable support for a significantly larger number of devices than pseudo-random sequences, i.e., a higher overloading factor without increasing the sequence length.

In spreading-based grant-free NOMA, each active device randomly and independently selects a spreading sequence from a predefined set \cite{cai2017modulation}. Therefore, the key research challenge is to correctly detect the spreading sequences of the active devices, also known as a multi-user detection (MUD) problem \cite{docomo2016uplink}. In this regard, identifying the total number of active devices, also known as the active user detection (AUD) sub-problem, and the accuracy of correctly identified active devices, which is the active user support set sub-problem, play a key role. This research challenge is addressed in this work.

\subsection{Related Work}
The quality of the active user support set devised from AUD directly impacts the performance of MUD. In many practical IoT use cases, while the total number of devices is large, only a small percentage of the total devices may be active in a given time frame \cite{hong2014sparsity, Mehmood2017, salari2022noma}. Using this inherent sparsity of IoT devices, the AUD problem can be readily formulated as a sparse recovery problem, which can be solved using compressed sensing (CS) \cite{tropp2007signal, dai2009subspace} or machine learning (ML) \cite{shahab2020grant}. Considering the inherent sparsity and the sporadic device activity, it is then crucial to correctly model the activity pattern of devices over a time frame. The activity pattern of devices over a given time frame, whether independent or temporally correlated, greatly impacts the performance of MUD.

In the literature, as summarised in Table~\ref{LitTab}, the frame-wise sparsity and burst-sparsity are two prominent models for device activity patterns adopted for both CS-based and ML-based MUD. In the frame-wise sparsity model, the activity and inactivity of devices remain constant over an entire data frame, i.e., the temporal correlation of device activity patterns is fixed. In the burst-sparsity model, devices generally transmit their data in multiple adjacent time slots with a high probability. In many IoT applications where devices are deployed to detect an event of interest \cite{9057670}, and device data is sporadic in time \cite{dawy2016toward, chen2020massive}, e.g., smart meters and environmental monitoring, IoT devices transmit their data in the form of data bursts due to the size of their payload. However, some IoT devices can still complete their transmission in a single time slot. Thus, the burst sparsity model is regarded as a more practical model since it provides a balance between the frame-wise sparsity (a slightly impractical model since the temporal correlation of the device activity patterns is fixed) and the completely random transmission model (also slightly impractical as IoT traffic is not entirely random and consists of data bursts and traffic patterns).

\textit{CS-based solutions with frame-wise sparsity model:} Many works have considered CS-based solutions for the MUD problem in spreading based grant-free NOMA with frame-wise sparsity \cite{abebe2016iterative, cirik2017multi, du2018joint, yu2019multiuser, wu2021joint}. In \cite{abebe2016iterative}, the frame-wise joint sparsity model is exploited to achieve better performance of device detection using an iterative order recursive least square (IORLS) algorithm based on the orthogonal matching pursuit (OMP) algorithm. However, the authors considered prior knowledge of device sparsity level at the AP, which is typically unknown in practical scenarios. In \cite{cirik2017multi}, the authors proposed the alternative-direction-method-of-multipliers-(ADMM)-based CS to show improvement in the device detection performance using a partial active device set as prior knowledge. However, obtaining the prior information on either the sparsity level, equivalent channel matrices, or both in practical systems is difficult. In \cite{du2018joint}, the device detection problem was modelled as a multiple measurement vector (MMV) problem, and a block sparsity adaptive subspace pursuit (BSASP) algorithm was used to solve it. However, pilot symbols are transmitted before every data packet, which leads to a significant system overhead. Similarly, the authors in \cite{yu2019multiuser, wu2021joint} developed greedy algorithms for joint device activity and data detection. However, these algorithms assume complete channel gain knowledge at the AP or pilot symbols for channel estimation.

\textit{CS-based solutions with burst sparsity model:} Some recent works have considered CS-based solutions for the MUD problem in spreading-based grant-free NOMA with burst sparsity \cite{wang2016dynamic, du2017efficient, cui2020side}. In \cite{wang2016dynamic}, a dynamic CS-based multi-device detection was proposed, which utilised the temporal correlation between device transmissions in the previous frame to achieve the performance gain. This algorithm was developed based on the assumption that the device sparsity level is known, which requires a training stage to learn such information accurately. Alternatively, the prior-information aided adaptive subspace pursuit (PIAASP) algorithm was proposed in \cite{du2017efficient}, which utilised the prior support according to the additional quality information (the number of common support sets shared in time slots). However, the preceding work is heavily dependent on the inertia of device support; thus, it is unsuitable when the active device support varies rapidly in adjacent time slots, as is often the case in practice. Similarly, the authors in \cite{cui2020side} proposed an algorithm to take advantage of the temporal correlation, where the frame is divided into subframes. Each subframe contains adjacent time slots and considers the active and inactive devices sharing common support in all the time slots. Also exploiting the temporal correlation, the authors in \cite{9522070} used $\ell_{2,1}$ minimisation to jointly detect the user activity and data.

\begin{table*}
\centering
\caption{Comparison of this work with recent related works in grant-free NOMA.}
\label{LitTab}
\resizebox{\linewidth}{!}{%
\begin{tabular}{|l|c|l|l|l|l|l|} 
\hline
\multicolumn{1}{|c|}{\textbf{References}} & \textbf{Strategy}                              & \multicolumn{1}{c|}{\textbf{Sparsity Model}}                                              & \multicolumn{1}{c|}{\textbf{Description}}                                                                                                                                                           & \multicolumn{1}{c|}{\textbf{Spreading Sequence}}                                              & \multicolumn{1}{c|}{\begin{tabular}[c]{@{}c@{}}\textbf{Overloading}\\\textbf{ Factor (\%)}\end{tabular}} & \multicolumn{1}{c|}{\begin{tabular}[c]{@{}c@{}}\textbf{\# of Active }\\\textbf{ Devices, $S$}\end{tabular}}  \\ 
\hline
\cite{abebe2016iterative} — 2016          & \multirow{10}{*}{\rotatebox{90}{Compressed sensing \quad\quad\quad\quad\quad\quad\quad\quad\quad\quad}} & \begin{tabular}[c]{@{}l@{}}Frame-wise \\ joint sparsity\end{tabular}                      & \begin{tabular}[c]{@{}l@{}}Low complexity joint device activity and\\ data detection using OMP based IORLS\\ algorithm.\end{tabular}                                                                & \begin{tabular}[c]{@{}l@{}}Binary spreading based\\ on pseudo-random noise\end{tabular}       & 200                                                                                                      & 10-20                                                                                                        \\ 
\cline{1-1}\cline{3-7}
\cite{cirik2017multi} — 2017              &                                                & \begin{tabular}[c]{@{}l@{}}Frame-wise\\ joint sparsity\end{tabular}                       & \begin{tabular}[c]{@{}l@{}}ADMM-based joint device activity and data\\ detection using partial active device set as\\ prior knowledge.\end{tabular}                                                 & \begin{tabular}[c]{@{}l@{}}Binary spreading based\\ on pseudo-random noise\end{tabular}       & 150                                                                                                      & 12                                                                                                           \\ 
\cline{1-1}\cline{3-7}
\cite{du2018joint} — 2018                 &                                                & \begin{tabular}[c]{@{}l@{}}Frame-wise \\ joint sparsity\end{tabular}                      & \begin{tabular}[c]{@{}l@{}}Joint device activity and data detection as an\\ MMV problem, then solved using BSASP\\ algorithm.\end{tabular}                                                          & \begin{tabular}[c]{@{}l@{}}Binary spreading based\\ on pseudo-random noise\end{tabular}       & 400                                                                                                      & 20                                                                                                           \\ 
\cline{1-1}\cline{3-7}
\cite{yu2019multiuser} — 2019             &                                                & \begin{tabular}[c]{@{}l@{}}Frame-wise \\ joint sparsity\end{tabular}                      & \begin{tabular}[c]{@{}l@{}}Joint device activity and data detection using\\ greedy algorithm.\end{tabular}                                                                                          & \begin{tabular}[c]{@{}l@{}}Randomly generated\\ bipolar sequence\end{tabular}                 & 200                                                                                                      & 20                                                                                                           \\ 
\cline{1-1}\cline{3-7}
\cite{wu2021joint} — 2021                 &                                                & \begin{tabular}[c]{@{}l@{}}Frame-wise \\ joint sparsity\end{tabular}                      & \begin{tabular}[c]{@{}l@{}}Joint user activity and data detection using \\ greedy algorithm with pilot symbols.\end{tabular}                                                                        & \begin{tabular}[c]{@{}l@{}}Binary spreading based\\ on pseudo-random noise\end{tabular}       & 200                                                                                                      & 18-20                                                                                                        \\ 
\cline{1-1}\cline{3-7}
\cite{wang2016dynamic} — 2016             &                                                & Burst-sparsity                                                                            & \begin{tabular}[c]{@{}l@{}}Joint device activity and data detection using\\ greedy algorithm with pilot symbols.\end{tabular}                                                                       & \begin{tabular}[c]{@{}l@{}}Binary spreading based\\ on pseudo-random noise\end{tabular}       & 200                                                                                                      & 20                                                                                                           \\ 
\cline{1-1}\cline{3-7}
\cite{du2017efficient} — 2017             &                                                & Burst-sparsity                                                                            & \begin{tabular}[c]{@{}l@{}}Joint device activity and data detection using\\ prior-information support set in continuous-\\ time slots.\end{tabular}                                                 & \begin{tabular}[c]{@{}l@{}}Binary spreading based\\ on pseudo-random noise\end{tabular}       & 200                                                                                                      & 18-20                                                                                                        \\ 
\cline{1-1}\cline{3-7}
\cite{du2018block} — 2018                 &                                                & Block-sparsity                                                                            & \begin{tabular}[c]{@{}l@{}}Joint device activity and data detection by\\ formulating a block-CS based signal\\ recovery framework.\end{tabular}                                                     & \begin{tabular}[c]{@{}l@{}}Binary spreading based\\ on pseudo-random noise\end{tabular}       & 200                                                                                                      & 20                                                                                                           \\ 
\cline{1-1}\cline{3-7}
\cite{cui2020side} — 2020                 &                                                & \begin{tabular}[c]{@{}l@{}}Dynamic frame-\\ wise and burst-\\ sparsity model\end{tabular} & \begin{tabular}[c]{@{}l@{}}Joint device activity and data detection by\\ exploiting temporal correlation of (in)active\\ devices in continuous time slots for signal\\ reconstruction.\end{tabular} & \begin{tabular}[c]{@{}l@{}}Binary spreading derived\\ from i.i.d Gaussian source\end{tabular} & 200                                                                                                      & 20                                                                                                           \\ 
\cline{1-1}\cline{3-7}
\cite{9522070} — 2022~                    &                                                & Burst-sparsity                                                                            & \begin{tabular}[c]{@{}l@{}}Joint device activity and data detection by\\utilising temporal correlation of devices\\using weighted~$\ell_{2,1}$ minimisation.\end{tabular}                           & \begin{tabular}[c]{@{}l@{}}Binary spreading based\\on pseudo-random noise\end{tabular}        & 200                                                                                                      & 18                                                                                                           \\ 
\hline
\cite{kim2020deep} — 2020                 & \multirow{7}{*}{\rotatebox{90}{Machine learning \quad\quad\quad\quad\quad}}    & \begin{tabular}[c]{@{}l@{}}Frame-wise \\ joint sparsity\end{tabular}                      & \begin{tabular}[c]{@{}l@{}}Active device detection using vanilla DNN\\ with no temporal correlation.\end{tabular}                                                                                   & Low density spreading                                                                         & 143                                                                                                      & 4                                                                                                            \\ 
\cline{1-1}\cline{3-7}
\cite{sivalingam2021deep} — 2021          &                                                & \begin{tabular}[c]{@{}l@{}}Frame-wise \\ joint sparsity\end{tabular}                      & Active device detection using vanilla DNN.                                                                                                                                                          & \begin{tabular}[c]{@{}l@{}}Complex spreading \\ sequence\end{tabular}                         & 200                                                                                                      & 1-2                                                                                                          \\ 
\cline{1-1}\cline{3-7}
\cite{miao2020grant} — 2020               &                                                & \begin{tabular}[c]{@{}l@{}}Partial sparsity\\with predefined\\pattern\end{tabular}        & \begin{tabular}[c]{@{}l@{}}Joint device activity and data detection using\\ LSTM with partial temporal correlation.\end{tabular}                                                                    & \begin{tabular}[c]{@{}l@{}}Binary spreading based\\ on pseudo-random noise\end{tabular}       & 200                                                                                                      & 1-8                                                                                                          \\ 
\cline{1-1}\cline{3-7}
\cite{zou2021joint} — 2021                &                                                & \begin{tabular}[c]{@{}l@{}}Frame-wise \\ joint sparsity\end{tabular}                      & \begin{tabular}[c]{@{}l@{}}Joint device activity and data detection using\\ generative network.\end{tabular}                                                                                        & \begin{tabular}[c]{@{}l@{}}Binary spreading based\\ on pseudo-random noise\end{tabular}       & 200                                                                                                      & 40                                                                                                           \\ 
\cline{1-1}\cline{3-7}
\cite{emir2021deepmud} — 2021             &                                                & \begin{tabular}[c]{@{}l@{}}Frame-wise \\ joint sparsity\end{tabular}                      & \begin{tabular}[c]{@{}l@{}}User detection using vanilla DNN\\ utilising pilot-aided power-domain NOMA.\end{tabular}                                                                                 & None                                                                                          & 9-25                                                                                                     & 4-6                                                                                                          \\ 
\cline{1-1}\cline{3-7}
\cite{9484069} — 2022                     &                                                & \begin{tabular}[c]{@{}l@{}}Frame-wise\\joint sparsity\end{tabular}                        & \begin{tabular}[c]{@{}l@{}}Joint device activity and data detection in\\ pilot-aided power-domain NOMA.\end{tabular}                                                              & None                                                                                          & 312                                                                                                      & 10                                                                                                           \\ 
\cline{1-1}\cline{3-7}
This work                                 &                                                & Burst-sparsity                                                                            & \begin{tabular}[c]{@{}l@{}}Joint user and data detection of temporally\\ correlated devices by utilising attention-based\\ BiLSTM network.\end{tabular}                                             & \begin{tabular}[c]{@{}l@{}}Complex spreading \\ sequence\end{tabular}                         & 200                                                                                                      & 20-40                                                                                                        \\
\hline
\end{tabular}
}
\end{table*}

\textit{ML-based solutions:} Recent works have adopted ML and demonstrated higher detection accuracy than conventional iterative algorithms \cite{kim2020deep, sivalingam2021deep, miao2020grant, zou2021joint, rahman2022multi, emir2021deepmud}. The authors in \cite{kim2020deep} and \cite{sivalingam2021deep} considered pseudo-random noise-based and complex spreading sequences, respectively, and proposed deep neural networks (DNN) for active user detection (D-AUD) in a grant-free NOMA system by using the received signal as the input to the DNN. However, since the preceding works utilised a vanilla DNN for this purpose, the temporal activity of the devices cannot be taken advantage of, leaving room for improvement. To tackle this, the authors in \cite{miao2020grant} utilised a long short-term memory (LSTM) network to predict the activity of the devices based on their activation history. However, the dependence of LSTM on the previous activation history of devices makes the overall system prone to misclassification since the activation history is vaguely modelled. Adopting a different approach, the authors in \cite{zou2021joint} considered the use-case of generative networks to tackle the issue of detecting devices in different overloading factors with a single trained model. However, this work did not take the temporal correlation of device activity patterns into account. The authors in \cite{emir2021deepmud} provided a somewhat different approach by utilising power-domain NOMA instead of code-domain NOMA as the multiple access signature. However, the system faces extreme degradation due to this choice as the number of active devices in the cell increases. Furthermore, pilot symbols are included after every data symbol, drastically increasing the system's overhead. In a similar fashion, the authors in \cite{rahman2022multi} utilised a bi-directional deep neural network for detection in a two-user power domain NOMA scenario. However, this differs from the grant-free NOMA scenario considered in this work since it does not use spreading-based signatures and assumes the connection of devices using prior access procedures. Similarly, the authors in \cite{9484069} assigned nonorthogonal pilots to devices for transmission, leading to a larger system overhead as the number of devices increases. Thus, a more resilient approach is required in the context of deep learning, which can exploit the temporal correlation of active devices in the adjacent time slots whilst providing accurate detection of devices. 

\subsection{Contributions}
As evident from Table 1, prior works on MUD in grant-free NOMA based on compressed sensing typically considered known user sparsity level or knowledge of the channel. Most works based on ML focused on the AUD problem with the frame-wise sparsity model, and the complete MUD problem was not considered. To the best of our knowledge, no prior work has fully addressed the MUD problem, along with AUD and active device support set identification subproblems, in spreading-based grant-free NOMA with burst sparsity model using ML when AP has no prior knowledge of user sparsity level and/or channel state.

In this paper, we consider a complex spreading sequences-based grant-free NOMA scenario, where multiple devices communicate with the AP simultaneously in the uplink following a burst-sparsity model. To address the AUD problem, we design an attention-based bidirectional LSTM (BiLSTM) network, which aims to create a mapping function between the superimposed received signal at the AP and the indices of active devices in the transmit signal. The proposed framework does not require active user sparsity or channel state knowledge to carry out AUD. Using the estimated active user support set, we then design a MUD framework to find the user sparsity and carry out blind data detection at the AP. The main contributions of this work are as follows: 
\begin{itemize}
    \item We design a BiLSTM network with an attention mechanism to carry out AUD. The BiLSTM network utilises two LSTM networks conjunctionally in opposite temporal directions. The attention mechanism exploits the temporal correlation in the active user set and facilitates the BiLSTM network by providing context to the important activation history of the active devices. By training the network in the offline stage, the proposed network maps the superimposed received signal and the active user support set, detecting a larger number of active devices with higher accuracy. 
    
    \item By detecting the active user support set using our proposed BiLSTM network, we then provide a framework to carry out blind data detection at the AP \cite{ameur2020power} without the need for explicit channel training. Using the estimated active user support set and complex spreading sequences, a blind minimum mean square error (MMSE) weight is obtained, from which the received signal is reconstructed without the explicit need for statistical channel information.
    
    \item Compared to the benchmark OMP scheme, our results show an improvement of around $30\%$ when detecting the number of active devices and an improvement of around $29\%$ when identifying the active device support set. Additionally, the proposed network achieves a gain of around $2.3$ dB in bit-error-rate (BER) compared to the OMP scheme.
    
    \item Compared to the ML-aided LSTM-based CS scheme, our results show an improvement of around $10\%$ when detecting the number of active devices and an improvement of around $6\%$ when identifying the active device support set. Additionally, the proposed network achieves a gain of around $0.9$ dB in BER compared to the LSTM-based CS scheme. The computational complexity of the proposed network increases only marginally as the number of active devices increases.
    
\end{itemize}

\subsection{Paper Organisation and Notations} The rest of this paper is organised as follows. In Section II, we present the system model and MUD problem. Section III describes the proposed attention-based BiLSTM scheme and describes the neural network's architecture. Section IV discusses the network's training details and complexity analysis. In Section V, we present the simulation results to verify the performance gain of the proposed technique. Finally, Section VI concludes the paper.

\begin{table}[t]
\caption{Important symbols used in this work.}
\label{tab:dimen}
\resizebox{0.9\linewidth}{!}{%
\begin{tabular}{|c|l|c|}
\hline
Variable                  & Description                                      & Dimension              \\ \hline
$K$                       & Total number of IoT devices                      & $1 \times 1$                    \\ \hline
$N$                       & Total subcarriers                                & $1 \times 1$                    \\ \hline
$S$                       & Active number of IoT devices                     & $1 \times 1$                  \\ \hline
$J$                       & Number of time slots                             & $1 \times 1$                    \\ \hline
$\mathbf{c}$              & Spreading sequence                               & $N \times 1$           \\ \hline
$\mathbf{g}$              & Channel                                          & $N \times 1$           \\ \hline
$\mathbf{v}$              & Synthesis of transmit symbol and channel         & $NK \times 1$          \\ \hline
$\tilde{\mathbf{v}}$      & Stacked synthesis of transmit symbol and channel & $NJK \times 1$         \\ \hline
$\mathbf{x}$              & Sparse transmit signal                           & $NJK \times 1$         \\ \hline
$\mathbf{y}$              & Received signal                                  & $NJ \times 1$          \\ \hline
$\tilde{\mathbf{y}}$      & Stacked received signal                          & $NJ \times 1$          \\ \hline
$\acute{\mathbf{y}}$      & Transformed sparse signal                        & $K \times 1$           \\ \hline
$\hat{\boldsymbol{\chi}}$ & Bits of reconstructed sparse signal              & $K \times 1$           \\ \hline
$\mathbf{\Gamma}$         & Active device support set                        & $K \times 1$           \\ \hline
$\hat{\mathbf{\Upsilon}}$ & Estimated active device support set              & $K \times 1$           \\ \hline
$\mathbf{z}^\text{out}$   & Network output                                   & $K \times 1$           \\ \hline
$\mathbf{b}^\text{out}$   & Output bias                                      & $K \times 1$           \\ \hline
$\hat{\mathbf{y}}_t$      & Initial BiLSTM input                             & $2NJ \times 1$         \\ \hline
$\mathbf{z}^\text{in}_t$  & Initial BiLSTM update                            & $\alpha \times 1$      \\ \hline
$\mathbf{b}^\text{in}_z$  & Initial BiLSTM bias                              & $\alpha \times 1$      \\ \hline
$\mathbf{b}$              & Hidden layer bias                                & $\alpha \times 1$      \\ \hline
$\mathbf{W}^\text{in}_z$  & Initial BiLSTM weight                            & $\alpha \times 2NJ$    \\ \hline
$\mathbf{W}$              & Hidden layer weight                              & $\alpha \times \alpha$ \\ \hline
$\mathbf{W}^\text{out}$   & Output weight                                    & $K \times \alpha$      \\ \hline
$\mathbf{C}$              & Codebook matrix                                  & $N \times NK$          \\ \hline
$\xi$                     & Rearranged codebook matrix                       & $NJ \times NJK$        \\ \hline
\end{tabular}%
}
\end{table}

We use the following notations in this paper. Lower and upper case boldface letters are used for vectors and matrices, respectively. The transpose of a vector $\mathbf{a}$ is $\mathbf{a}^T$. The exponential function is calculated as $e^{(\cdot)}$, where $e$ is the base of the natural logarithm. The norm is denoted by $\left|\left| \cdot \right|\right|$. $\mathbb{R}^{x \times y}$ and $\mathbb{C}^{x \times y}$ denotes the real and complex valued space of size $x \times y$ respectively. $\mathrm{diag} ( \cdot )$ denotes the diagonal operation, $\odot$ denotes Hadamard product, whereas $\oplus$ denotes the symmetric difference. $\Re(\cdot)$ and $\Im(\cdot)$ denote the real and imaginary parts of a complex number respectively. The gradient differential operator is denoted by $\nabla$. $\bar{\mathbf{z}}$, $\tilde{\mathbf{z}}$, $\hat{\mathbf{z}}$, $\check{\mathbf{z}}$, and $\breve{\mathbf{z}}$ represent the output states of the proposed network at their respective intermediate stages. Table~\ref{tab:dimen} summarizes the important symbols used in this work, including the dimensions of vectors and matrices.

\section{System Model}
\begin{figure}[t]
\centering
\includegraphics[scale=0.5]{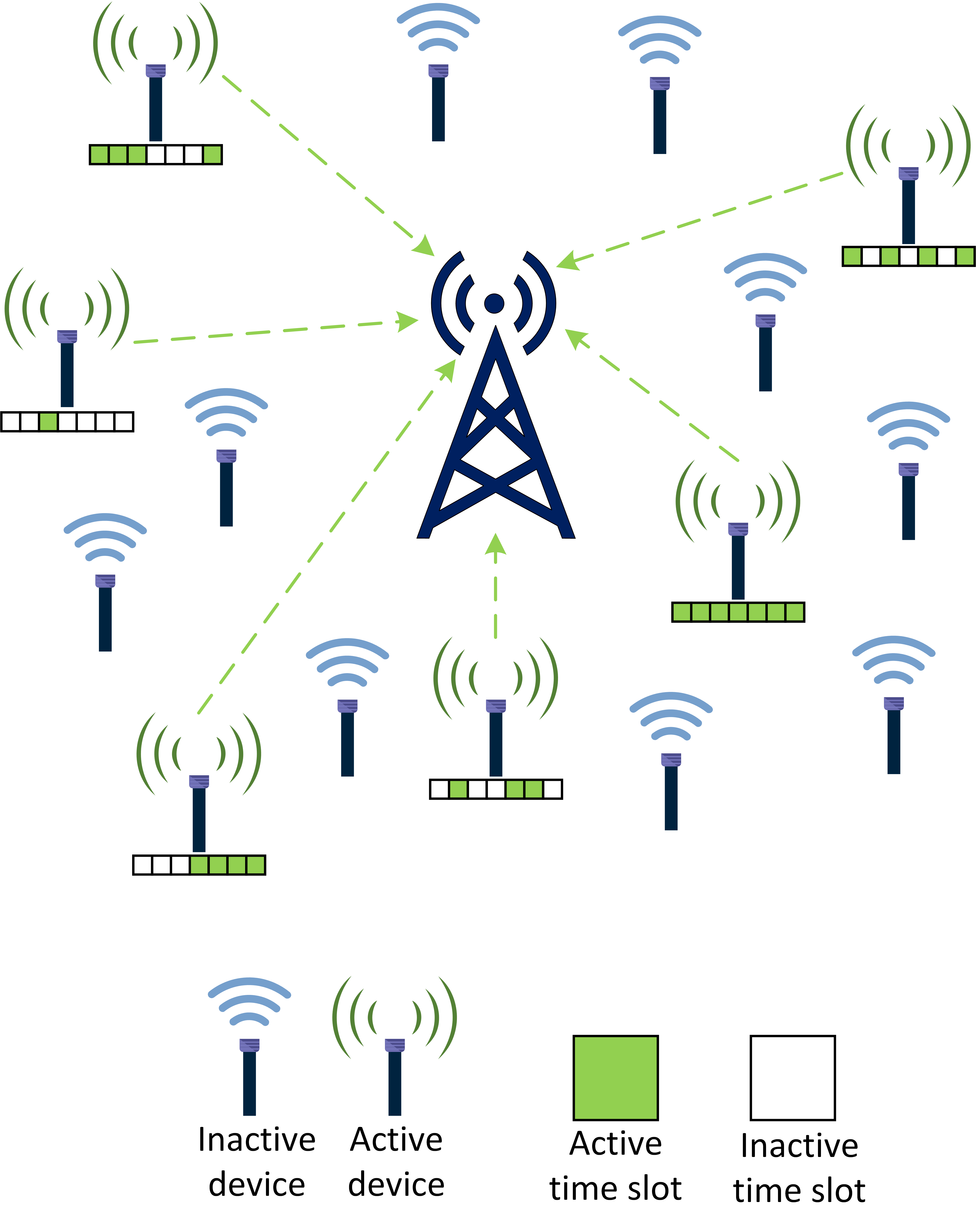} 
\caption{Illustration of a typical uplink grant-free NOMA system.}
\label{SystemMod}
\end{figure}

We consider a spreading-based uplink grant-free NOMA system comprising of an AP and $K$ IoT devices, as shown in Fig.~\ref{SystemMod}. Without loss of generality, all devices and the AP are assumed to be equipped with a single antenna. We consider an overloaded system where the number of resource blocks $N$ is less than the number of IoT devices, i.e., $N < K$. During transmission, a subset of the $K$ devices sporadically and randomly become active when they have data to transmit. We adopt the burst-sparsity model in this work, i.e., some transmissions continue for several consecutive time slots while others last for one-time slot only \cite{wang2016dynamic, du2017efficient, cui2020side}.

\subsection{Signal Model}
Considering an arbitrary symbol interval, an active device transmits its complex modulated signal towards the AP, which are independent random variables drawn from standard symmetric discrete constellation set $M$. For inactive devices, their transmit symbol is equal to zero. In this work, we consider that the device symbols are spread with a family of short complex-valued spreading sequences with low cross-correlation values \cite{yuan2016multi}. These short complex-valued spreading sequences can be generated naturally based on the binary sequence elements. For instance, for $M = 3$, each element of the complex spreading sequence is taken from the set $\{ -1,0, 1, -1+i, i, 1+ i, -1-i, -i, 1-i \}$\cite{yuan2016multi,sivalingam2021deep}.

After modulation, the symbol $s_k$ from the $k^{\text{th}}$ device is spread onto a spreading sequence $\mathbf{c}_k = [c_{1k}, c_{2k}, \hdots, c_{Nk}]^T \in \mathbb{C}^{N \times 1}$ which is randomly and independently selected from a pre-defined set.
The received signal $\mathbf{y}$ at the AP is the superposition of all signals, given as 
\begin{equation} \label{eqmain}
        \mathbf{y}  =	\sum_{k=1}^{K}   \mathrm{diag}(\mathbf{c}_k) \mathbf{g}_k  s_k   +   \mathbf{w} \; = \mathbf{C} \mathbf{v}  +   \mathbf{w}
\end{equation}
where $\mathbf{g}_k = [g_{1k}, g_{2k}, \hdots , g_{Nk}]^T \in \mathbb{C}^{N \times 1}$ denotes the channel vector between the AP and the $k^{\text{th}}$ device over $N$ sub-carriers, and $\mathbf{w} \sim \mathcal{CN}(0 , \sigma^2 \mathbf{I})$ represents the complex  Gaussian noise vector. 
Moreover, $\mathbf{C} = [\mathrm{diag}(\mathbf{c}_1),   \mathrm{diag}(\mathbf{c}_2), \hdots, \mathrm{diag}(\mathbf{c}_K)] \in \mathbb{C}^{N \times NK}$ refers to the codebook matrix of all devices, and $\mathbf{v} = [ \mathbf{v}_1^T, \mathbf{v}_2^T, \hdots, \mathbf{v}_K^T]^T = [(s_1\mathbf{g}_1)^T, (s_2\mathbf{g}_2)^T, \hdots, (s_K\mathbf{g}_K)^T]^T \in \mathbb{C}^{NK \times 1}$ is the synthesis of the transmit symbols and channel vectors.

\subsection{Consecutive-Time Slot Dynamic Model} \label{sec:consec}
Exploiting the sparsity in the data transmission (i.e., only a subset of devices wake up to transmit) and the temporal correlation of the device activity pattern (i.e., data transmission is bursty in general), we can formulate the vector $\mathbf{v}$ as a sparse vector and extend our system model in (\ref{eqmain}) to a continuous-time slot model.

The idea is to utilise the bursty nature of $\tilde{\mathbf{v}} = \left[ \mathbf{v}^{[1]}, \mathbf{v}^{[2]}, \hdots , \mathbf{v}^{[J]} \right]^{T} \in \mathbb{C}^{NJK \times 1}$ where $\mathbf{v}^{[j]}$ is the signal at the $j$-th time slot, to retrieve it from the received signals $\tilde{\mathbf{y}} = \left[ \mathbf{y}^{[1]}, \mathbf{y}^{[2]}, \hdots , \mathbf{y}^{[J]} \right]^{T} \in \mathbb{C}^{NJ \times 1}$, in the $J$ successive time slots. This formulation helps in capturing the temporal correlation of the active devices by detecting the transmit signals $\mathbf{v}$ in the continuous-time slots. The stacked received signal vector $\tilde{\mathbf{y}}$ can be represented as
\begin{equation} \label{mat}
\tilde{\mathbf{y}} = 
\begin{bmatrix}
\mathbf{C}^{[1]} & 0 & \cdots & 0 \\
0 & \mathbf{C}^{[2]} & \cdots & 0 \\
\vdots  & \vdots  & \ddots & \vdots  \\
0 & 0 & \cdots &\mathbf{C}^{[J]}
\end{bmatrix}
\begin{bmatrix}
\mathbf{v}^{[1]}  \\
\mathbf{v}^{[2]}  \\
\vdots            \\
\mathbf{v}^{[J]}
\end{bmatrix}
+
\begin{bmatrix}
\mathbf{w}^{[1]}  \\
\mathbf{w}^{[2]}  \\
\vdots            \\
\mathbf{w}^{[J]}
\end{bmatrix},
\end{equation}
where $\mathbf{C}$ is the equivalent code-book matrix of all devices, which contains the complex spreading sequences of all $K$ devices, $\mathbf{v}$ is the composite of the transmitted symbol and channel vector, and $\mathbf{w}$ is the Gaussian noise vector.

The AP receives a multi-device vector $\tilde{\mathbf{y}}$ with no knowledge of the active transmitting devices or locations of the non-zero symbols. The active device support set $\mathbf{\Gamma}^{[j]}$ varies over different time slots considering the device's random transmission in a grant-free fashion. With this in mind, let $\mathbf{u}^{[j]} = [s_1, s_2, \hdots, s_K]^T$ correspond to the total devices in the $j$-th time slot. Then, the active device support set\footnote{The actual device support set $\mathbf{\Gamma}^{[j]}$ is utilised as ground truth to reduce the misclassification rate during network training only. In the online deployment, the proposed BiLSTM network is used to estimate the active device support set. This will be explained in the following sections.} $\mathbf{\Gamma}^{[j]}$ of the signal $\mathbf{x}^{[j]}$ in the $j$-th time slot is defined as \cite{du2017efficient}
\begin{equation} \label{eqSupportSet}
    \mathbf{\Gamma}^{[j]} = { k \; | \; \mathbf{u}_k^{[j]} \neq 0, 1 \leq k \leq K }.
\end{equation}
From this, the number of transmitting active devices is defined through the cardinality of the active device support set $\mathbf{\Gamma}^{[j]}$, given as \cite{du2017efficient}
\begin{equation}
    S^{[j]} = \left| \left| \mathbf{\Gamma}^{[j]} \right| \right|_0. 
\end{equation}

Since IoT traffic is not entirely random and often consists of data bursts and traffic patterns, in this work, we consider the burst-sparsity model where only a subset of active devices in the previous time slot also transmit in the next time slot. That is, only a subset of indices in $\mathbf{\Gamma}^{[j-1]}$ are present in $\mathbf{\Gamma}^{[j]}$. Therefore, to quantify the commonality of active devices transmitting in consecutive time slots, we define $\eta$ as the level of temporal correlation between the previous time slot $\mathbf{\Gamma}^{[j-1]}$ and the current time slot $\mathbf{\Gamma}^{[j]}$. It is given as
\begin{equation} \label{tempCorr}
    \eta = \frac { \left| \left|  \mathbf{\Gamma}^{[j-1]} \bigcap \mathbf{\Gamma}^{[j]} \right| \right|_0 } { \left| \left| \mathbf{\Gamma}^{[j]} \right| \right|_0 }.
\end{equation}

Note that in (\ref{tempCorr}), $\eta$ characterises the overlapping level of the active devices transmitting in consecutive time slots. For instance, when $\eta = 0.5$, half of the devices transmit in consecutive time slots $\geq 2$, whereas the remaining transmit only once during the whole process. In Section V, we will show how the variation of temporal correlation $\eta$ affects the overall system performance.

\subsection{Multi-User Detection Problem}
When multiple active devices communicate with the AP simultaneously in a grant-free manner, the first task for the AP is to detect the active devices that contributed to the received signal. Therefore, the identification of active devices leads to the problem of finding the support of the transmitted signal. 

In this regard, the rows in (\ref{mat}) can be rearranged. We also introduce an active device criterion $\delta \in {0,1}$, where $\delta=1$ and $\delta=0$ correspond to active and inactive devices, respectively \cite{kim2020deep}. Using this, the stacked received signal vector $\tilde{\mathbf{y}}$ can be written as
\begin{equation} \label{eqReceiveSig}
    \tilde{\mathbf{y}} = 
\begin{bmatrix}
\xi_1 & \cdots & \xi_K \\
\end{bmatrix}
\begin{bmatrix}
\delta_1 \mathbf{x}_1  \\
\vdots            \\
\delta_K \mathbf{x}_K  \\
\end{bmatrix}
+
\begin{bmatrix}
\mathbf{w}^{[1]}  \\
\vdots            \\
\mathbf{w}^{[J]}
\end{bmatrix}
 =   \xi \mathbf{x + w},
\end{equation}
where $\xi = [\xi_1, \xi_2, \hdots, \xi_K] \in \mathbb{C}^{NJ \times NJK}$, and $\mathbf{x} = [\delta_1 \mathbf{x}_1^T, \delta_2 \mathbf{x}_2^T, \hdots, \delta_K \mathbf{x}_K^T]^T \in \mathbb{C}^{NJK \times 1}$, such that for any $k^{\text{th}}$ device, $\mathbf{x}_k =  [(s_k^{[1]} \mathbf{g}_k^{[1]})^T, (s_k^{[2]} \mathbf{g}_k^{[2]})^T, \hdots, (s_k^{[J]} \mathbf{g}_k^{[J]})^T]^T $ and $\xi_k = [\mathrm{diag}(\mathbf{c}_k^{[1]}), \mathrm{diag}(\mathbf{c}_k^{[2]}),  \hdots, \mathrm{diag}(\mathbf{c}_k^{[J]})]$, respectively.
From (\ref{eqReceiveSig}), it is inferred that out of $K$, only a subset of devices, say $S$, are active. This means that the sparse vector $\mathbf{x}$ has $S$ nonzero blocks corresponding to the $S$ active devices. Therefore, $\tilde{\mathbf{y}}$ in \eqref{eqReceiveSig} can be represented as a linear combination of $S$ submatrices of $\xi_1, \hdots,  \xi_K$ perturbed by the noise \cite{kim2020deep}. We assume the codebook entries of $\xi$ are available at the AP \cite{shahab2020grant}. However, the AP does not know which spreading sequence is chosen by the different active devices. Thus, the AP needs to identify the sub-matrices $\xi_S$, which are analogous to $\xi$, by processing $\tilde{\mathbf{y}}$.

From this, the MUD problem becomes a 2-dimensional CS problem, which is common in the CS paradigm \cite{choi2017compressed}. With this in mind, the following  MUD problem is readily articulated as an active device support estimation problem, given as
\begin{equation} \label{eqprob}
     \mathbf{\Upsilon} = \underset{\mathbf{|\Upsilon|} = S}{\text{arg min }} \, \frac{1}{2}\left|\left|  \tilde{\mathbf{y}} - \xi_{\mathbf{\Upsilon}} \mathbf{x}_{\mathbf{\Upsilon}}  \right|\right|^2_2.
\end{equation}

This detection problem in (\ref{eqprob}) can be solved using classical CS approaches. The approaches based on exhaustive searches, such as $\ell_1$-minimisation \cite{chen1994basis}, provide theoretical performance gains but suffer from heavy computational complexity. The approaches based on greedy algorithms \cite{donoho2012sparse} have comparably lower complexity but result in a sub-optimal solution and require a larger number of measurements for signal recovery. The biggest drawback of conventional CS-based schemes is that they assume perfect knowledge of the channel and active device sparsity levels. Furthermore, the enormous computational complexity and the latency of iterative algorithms make them a practical solution only for a small number of active devices. When there is a larger number of active devices, the performance of conventional CS-based schemes degrades due to their sole dependence on the residual vector in each iteration\footnote{A nonzero submatrix of $\xi$ with an index chosen at the $i$-th iteration is given as $
    \epsilon = \underset{k=1,\cdots,K}{\text{arg max }} \frac{1}{2}\left|\left|  \xi_k^H \mathbf{r}^{i-1}  \right|\right|^2_2$,
where $\mathbf{r}^{i-1} = \mathbf{y} - \xi_{\Upsilon}^{i-1}\hat{\mathbf{x}}^{i-1}$ is the $i$-th residual vector and $\hat{\mathbf{x}}^{i-1} = \xi_{\Upsilon^{i-1}}^\dagger \mathbf{y}$ is an approximate of the transmitted signal $\mathbf{x}$ in the $(i − 1)$-th iteration. It is of understanding that the performance of active user support identification is influenced primarily by $\xi$, which is generated through the codebook $\mathbf{C}$, and residual vector $\mathbf{r}^{(\cdot)}$.}. Due to this, as the number of active devices increases, conventional CS-based schemes are not suitable solutions to facilitate grant-free communication. This motivates us to pursue a machine learning-aided solution presented in the next section.

\section{Deep Learning aided MUD}
To tackle the MUD problem in Section II-C, we propose a solution using deep learning. In essence, we aim to delineate a nonlinear mapping using deep learning to create a pattern between the stacked received signal $\tilde{\mathbf{y}}$ and the support of $\mathbf{x}$ and perform MUD at the AP. The resulting active device support estimation problem $\hat{\mathbf{\Upsilon}}$ is then defined as 
\begin{equation}
    \hat{\mathbf{\Upsilon}} = g(\tilde{\mathbf{y}}, \Theta),
\end{equation}
where $\Theta$ represents the weights and corresponding biases of the learning architecture.

\subsection{Learning Architecture}

\begin{figure*}[t] 
\centering
\includegraphics[scale=0.5]{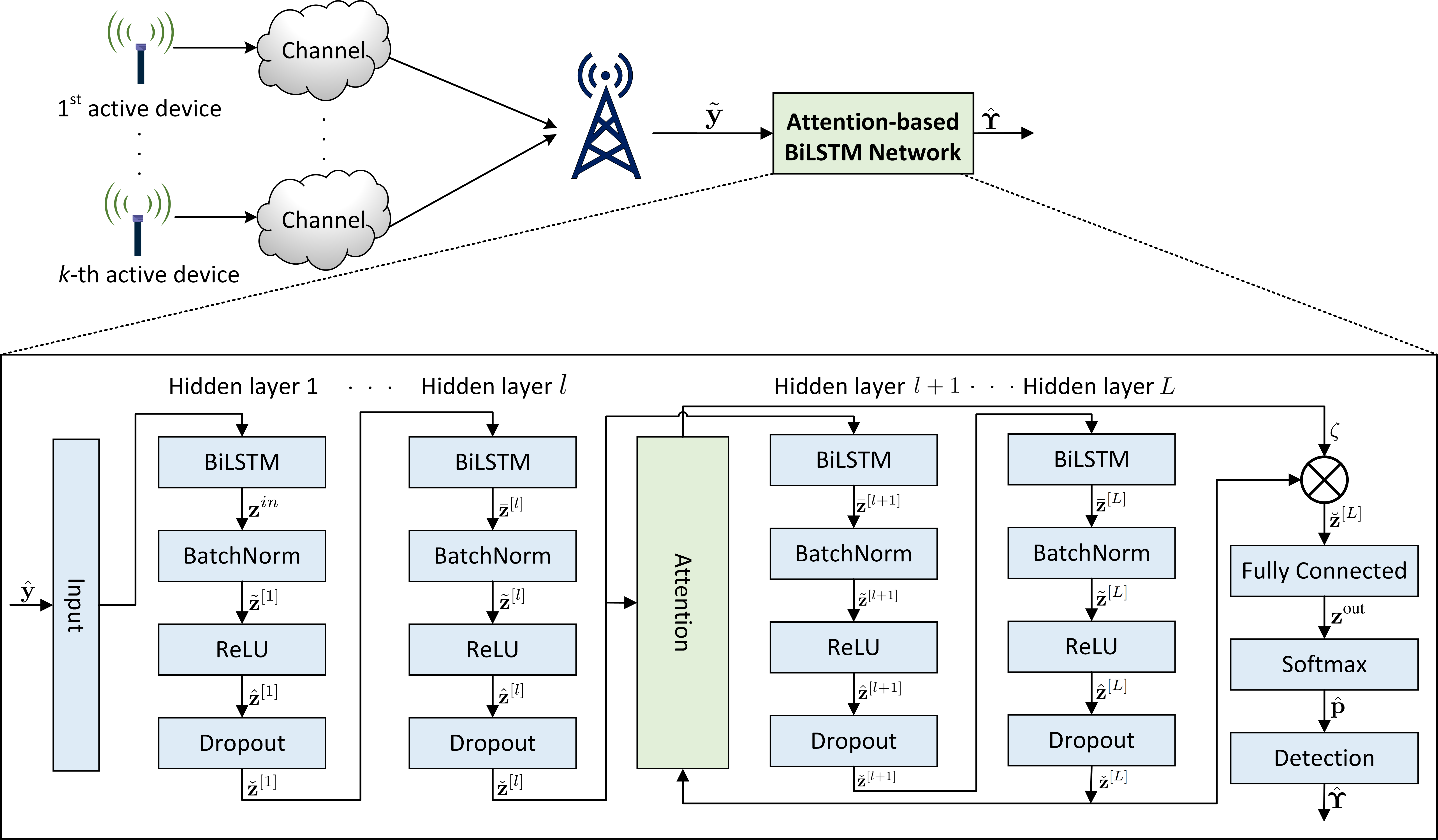} 
\caption{Detailed architecture and working of the proposed attention-based BiLSTM network.}
\label{figModel}
\end{figure*}

In this work, we adopt an attention-based BiLSTM network to solve the MUD problem, as illustrated in Fig.~\ref{figModel}. The attention mechanism is discussed in Section III-B, while in this section, we discuss the BiLSTM network. The motivation for adopting the BiLSTM network is as follows.

\indent 
Standard unidirectional LSTM networks undertake sequences in forwarding temporal order, ignoring future context. This is because unidirectional LSTM only preserves the information of the previous time steps since it has exclusively seen inputs from the past. On the other hand, BiLSTM networks take unidirectional LSTM networks one step further by setting up a second LSTM layer, where the gradients in the hidden connections flow in the opposite temporal direction. That is, BiLSTM runs the inputs in two ways, one from past to future (left to right, i.e., forward) and another one from future to past (right to left, i.e., backward). This gives BiLSTMs the ability to exploit more information, thereby simultaneously obtaining contextual features from forward and reverse temporal directions. In essence, more features from both directions are captured for mapping active devices transmitting in consecutive time slots. The LSTM in the reverse direction is calculated in the same fashion as the forward direction. Noticeably, since the direction is reversed, the time information is passed from future to past. 

For input $\tilde{\mathbf{y}_t}$ at the current time step $t$, the BiLSTM network calculation is given by
\begin{equation}
    \overrightarrow{\mathbf{h}_{f}}  = \sigma (\mathbf{W}_f \tilde{\mathbf{y}}_t  +  \mathbf{W}_f \mathbf{h}_{t-1} + \mathbf{b}_f),
\end{equation}
\begin{equation}
    \overleftarrow{\mathbf{h}_{r}}  = \sigma (\mathbf{W}_r \tilde{\mathbf{y}}_t  +  \mathbf{W}_r \mathbf{h}_{t+1} + \mathbf{b}_r),
\end{equation}
where $\sigma$ represents the activation function, $t-1$ and $t+1$ represent the forward and reverse direction time steps respectively, $\mathbf{h}_{t-1}$ and $\mathbf{h}_{t+1}$ represent the previous and next hidden states respectively, $\mathbf{W}_f$ and $\mathbf{W}_r$ represent the forward and reverse direction input weights respectively, and $\mathbf{b}_f$ and $\mathbf{b}_r$ represent the forward and reverse direction learnable bias parameter respectively. $\overrightarrow{\mathbf{h}_{f}}$ and $\overleftarrow{\mathbf{h}_{r}}$ represent the forward and reverse direction LSTM network outputs respectively. Finally, the output of the BiLSTM $\mathbf{z}_t$ is
\ifCLASSOPTIONonecolumn
\begin{equation} \label{eqUnitOutput}
    \mathbf{z}_t =  \sigma (\mathbf{W}_z \overrightarrow{\mathbf{h}_{f}}  \oplus \mathbf{W}_z \overleftarrow{\mathbf{h}_{r}} + \mathbf{b}_z) \;  =  \sigma (\mathbf{W}_z \tilde{\mathbf{h}}_{t}  + \mathbf{b}_z),    
\end{equation}
\else
\begin{equation} \label{eqUnitOutput}
    \mathbf{z}_t =  \sigma (\mathbf{W}_z \overrightarrow{\mathbf{h}_{f}}  \oplus \mathbf{W}_z \overleftarrow{\mathbf{h}_{r}} + \mathbf{b}_z) \;  =  \sigma (\mathbf{W}_z \tilde{\mathbf{h}}_{t}  + \mathbf{b}_z),    
\end{equation}
\fi
where $\mathbf{W}_z$ represents the BiLSTM output weights, $\mathbf{b}_z$ represents the BiLSTM output learnable bias parameter, and $\tilde{\mathbf{h}}_{t}$ is the concatenated hidden state of the forward and reverse direction LSTMs.

Fig.~\ref{figModel} shows the proposed attention-based BiLSTM network applied to our MUD problem. For each training iteration, we use $U$ training data copies $\tilde{\mathbf{y}}^{(1)}, \cdots, \tilde{\mathbf{y}}^{(U)}$. Next, since the stacked received signal $\tilde{\mathbf{y}}^{(u)}$ is a complex-valued modulated vector, we split the real and imaginary parts and use $\hat{\mathbf{y}} ^{(u)} = [ \Re( \tilde{y}^{(u)}_{1} ) \cdots \Re( \tilde{y}^{(u)}_{N} ),  \Im( \tilde{y}^{(u)}_{1} ) \cdots \Im( \tilde{y}^{(u)}_{N} )]$ as an input vector to the network. With this in mind, the unit output in (\ref{eqUnitOutput}) is substituted as
\begin{equation} \label{eqOutputVec}
    \mathbf{z}^{\text{in}, (u)}_t =  \sigma (\mathbf{W}^{\text{in}}_z \hat{\mathbf{y}} ^{(u)}_t + \mathbf{b}^{\text{in}}_z), \quad \text{for} \;  u = 1, \cdots, U,
\end{equation} 
where $\mathbf{W}^{\text{in}}_z \in \mathbb{R}^{\alpha \times 2NJ}$ is the initial weight, $\hat{\mathbf{y}} ^{(u)}_t \in \mathbb{R}^{2NJ \times 1}$ is the input vector, and $\mathbf{b}^{\text{in}}_z \in \mathbb{R}^{\alpha \times 1}$ is the input learnable bias term. 

In this work, we employ batch normalisation to help coordinate the update of multiple layers by standardising the inputs of each layer to have fixed means and variances. This is important because when active devices experience different wireless channels and transmit their data in a grant-free manner, the resulting stacked received signal $\tilde{\mathbf{y}}$ has substantial variations. These significant variations make it difficult for the network to learn the device activation pattern. By standardising the inputs of each layer, batch normalisation reduces the variations and helps to overcome this difficulty. Thus, the output vectors $U$ from (\ref{eqOutputVec}) are put together in the mini-batch $\mathbf{B} = [\mathbf{z}^{(1)}_t \cdots \mathbf{z}^{(U)}_t ]^T$. Once arranged in a mini-batch, these vectors are scaled and shifted using their respective hidden weights and batch normalised. The output for each element $z^{\text{in}, (u)}_{t,i}$ of the batch normalisation (BatchNorm) is given as
\begin{equation}
    \tilde{z}^{(u)}_{t,i} = \beta \left(  \frac { z^{\text{in}, (u)}_{t,i}  - \mu_{\mathbf{B}, t, i}}    {  \sqrt{\sigma^2_{\mathbf{B}, t, i}}  }         \right)  + \gamma, \quad \text{for} \;  i = 1, \cdots, \alpha,
\end{equation}
where $\mu_{\mathbf{B}, t, i} = \frac{1}{U} \sum_{u=1}^U z^{\text{in}, (u)}_{t,i}$ calculates the batch-wise mean, $\sigma^2_{\mathbf{B}, t, i} = \frac{1}{U} \sum_{u=1}^U (z^{\text{in}, (u)}_{t,i} - \mu_{\mathbf{B}, t, i})^2$ calculates the batch-wise variance, $\beta$ is used as a scaling parameter, $\gamma$ is used as the shifting parameter, and $\alpha$ represents the width of the hidden layers.

The proposed scheme learns to create a mapping function between the stacked received signal $\tilde{\mathbf{y}}$ and the current active device support set $\mathbf{\Gamma}$. However, the estimate of the current active device support set $\hat{\mathbf{\Upsilon}}$ is vastly agitated by the activation patterns of the neurons, which in turn are dependent on perturbations and precision errors. This issue is further compounded as the spreading sequences in the sensing codebook matrix $\xi$ are correlated. Accordingly, the estimate of the current active device support set $\hat{\mathbf{\Upsilon}}$ might not be accurate and will misclassify in the presence of random perturbations. In addition, when the device activity pattern is similar in consecutive time slots, the network is more prone to overfitting due to the unchanging device activation pattern. We use the ReLU activation function and dropout layer to address these issues. By using the ReLU activation function, the computed weights at every iteration are ranged, i.e., $f(x) = \text{max}(x, 0)$, which is then used as
\begin{equation}
    \hat{\mathbf{z}}^{(u)}_{t} = f(\tilde{\mathbf{z}}^{(u)}_{t}).
\end{equation}
In the dropout layer, the activated neurons in a hidden layer are randomly halted with a probability $\rho_{\text{drop}}$, given as
\begin{equation}
    \check{\mathbf{z}}^{(u)}_{t} = \hat{\mathbf{z}}^{(u)}_{t} \odot \mathbf{d}^{(u)}, \quad  d^{(u)}_i \sim  Bern(\rho_{\text{drop}})
\end{equation}
where $d^{(u)}_i$ is the $i$-th element of the dropout vector $\mathbf{d}^{(u)}$, and $\odot$ is the Hadamard product. $Bern(\rho_{\text{drop}})$ is the Bernoulli random variable which takes the value $0$ with the dropout probability $\rho_{\text{drop}}$ and $1$ with the probability $1 - \rho_{\text{drop}}$. The dropout mechanism deliberately makes the training process noisy by deactivating neurons randomly, forcing the remaining neurons to take more responsibility in creating a different path for the gradient flow. This random dilution of neurons provides rigorous circumstances where network layers co-adapt to rectify mistakes from prior layers, which helps create a more generalised network capable of estimating the current active device support set with more accuracy. Therefore, removing incoming and outgoing connections of the dropped neurons with a random probability systematically resolves the active device activation patterns' similarity among correlated support sets.

After the dropout layer, the output vector $\check{\mathbf{z}}$ makes its way through multiple hidden layers\footnote{For notational simplicity, in the proceeding sections, the training data index $u$ has been omitted.}. In subsequent, every hidden layer comprises the BiLSTM layer, a BatchNorm layer to reduce the variation of $\bar{\mathbf{z}}^{[l]}$, a ReLU activation function applied to $\tilde{\mathbf{z}}^{(u)}$ to determine whether the information $(\hat{z}^{[l]}_{1}, \hdots, \hat{z}^{[l]}_{\alpha} )$ generated by the hidden unit is activated or not, and finally a dropout layer to overcome overfitting of the network is applied (see Fig.~ \ref{figModel}). The output of the $l^{\text{th}}$ hidden layer's BiLSTM $\bar{\mathbf{z}}^{[l]}_t$ is
\begin{equation}
    \bar{\mathbf{z}}^{[l]}_t = \mathbf{W}^{[l]} \left( \sum_{i=1}^{l-1} \check{\mathbf{z}}^{[i]}_{t}      \right) + \mathbf{b}^{[l]},
\end{equation}
where $\mathbf{W}^{[l]} \in \mathbb{R}^{\alpha \times \alpha}$ and $\mathbf{b}^{[l]} \in \mathbb{R}^{\alpha \times 1}$ are the weight and bias in the $l^{\text{th}}$ hidden layer, respectively.

\subsection{Attention Mechanism}
\begin{figure}[t] 
\centering
\includegraphics[scale=0.43]{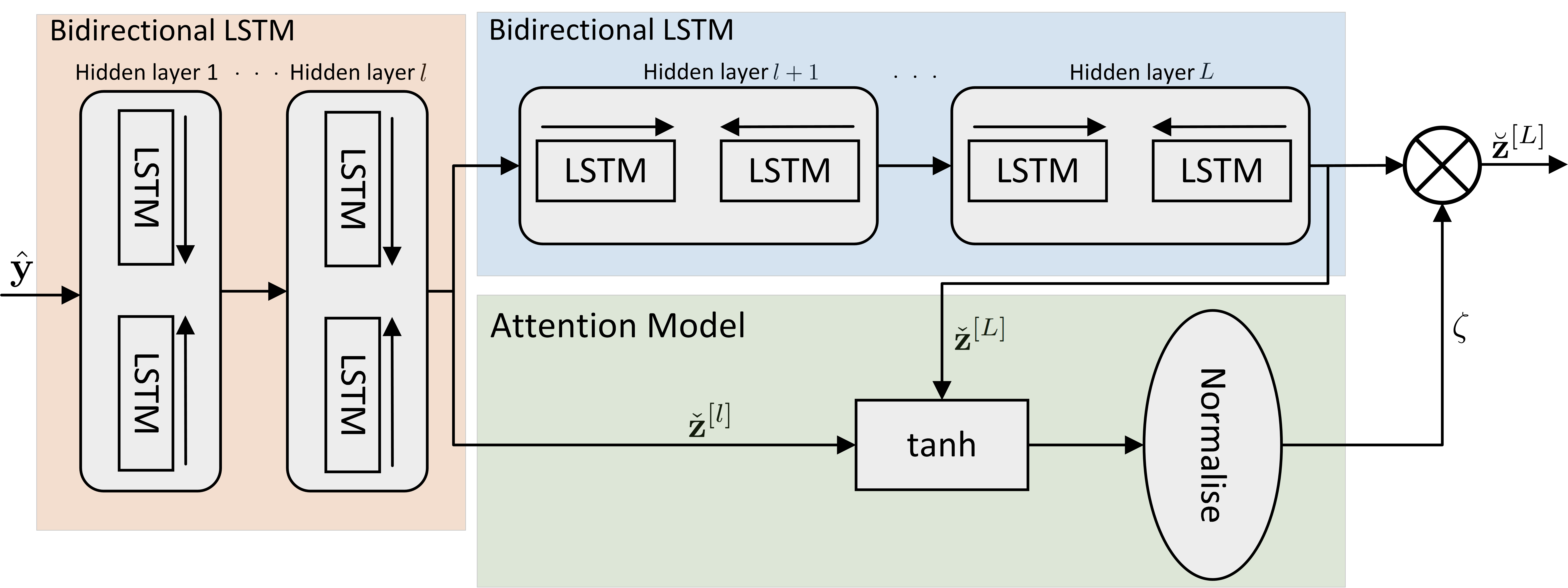} 
\caption{The proposed BiLSTM module with an attention mechanism.}
\label{figAtten}
\end{figure}

Fig.~\ref{figAtten} shows the working of the attention mechanism and its integration with the BiLSTM network architecture. The motivation for adopting the attention mechanism is twofold: (i) a neural network that creates a mapping function for the active device detection problem in (\ref{eqprob}) by analyzing the whole input at every step ignores the temporal correlation of the device activity pattern and (ii) with the increasing number of active devices, it becomes difficult for a neural network to learn its activation pattern due to its inherent sequential path architecture, causing problems such as vanishing and exploding gradients \cite{goodfellow2016deep}. 

The attention mechanism allows the neural network to apply context to specific parts of the data at every time step. That is, instead of finding the active devices in all of the input vectors altogether, a neural network with an attention mechanism breaks down the  data, applies a contextual vector to it, and then gives a score to the parts where active devices are present and transmitting consecutively. This mechanism brings additional temporal-based reasoning into the overall architecture for active device detection, helping the neural network load more active devices for detection.

The output of the BiLSTM network is computed as a weighted summation of the output of the BiLSTM network $\check{\mathbf{z}}^{(L)}_t$ at time step $t$ as
\begin{equation}
    \breve{\mathbf{z}}^{[L]}_t = \sum^{K+1}_{k = 1} \zeta_{k} \check{\mathbf{z}}^{[L]}_{t-(k-1)},
\end{equation}
where $\zeta_k$ is the temporal attention value at time step $t-(k-1)$, computed as
\begin{equation} \label{eqAttenSum}
    \zeta_{k} = \frac   {    e^{s_k}   }        {  \sum_{k=1}^{K} e^{s_k} },
\end{equation}
where the scores $\mathbf{s} = [s_1 \cdots s_K ]^T$ indicate the repeated activation pattern of active devices in the time slots, which is obtained as
\begin{equation} \label{eqAttenScore}
    s_{k} = W_{\text{rel}} \ \text{tanh}  (W_a \check{\mathbf{z}}^{[l]}_{t} + Z_a \check{\mathbf{z}}^{[L]}_{t} + b_a ).
\end{equation}
where $\check{\mathbf{z}}^{[l]}$ is the output of the previous hidden layers, and $W_a$ and $Z_a$ are the attention learnable parameters that learn to project each context element and hidden state into a latent space and $W_{\text{rel}}$ denotes the relevance parameter \cite{8365878}.

Evidently from (\ref{eqAttenSum}) and (\ref{eqAttenScore}), at time step $t$, $\zeta$ depends on the input $\mathbf{\check{z}}^{(l)}_{t}$. Furthermore, $\zeta$ is also dependent on the hidden variables $\check{\mathbf{z}}^{[L]}_{t}$ in the previous and current time step $t$. The attention value $\zeta$ can also be regarded as activating the active device detection gate. That is, the amount of information flow into the BiLSTM network is controlled by setting the gates. With this in mind, the final prediction result is influenced by a larger activation value, which results in a larger flow of information. It should also be noted that the standard LSTM network cannot detect many active devices concerning the previous activation pattern due to the large memory overhead occurring. The BiLSTM network with an attention-based mechanism can capture device activation patterns in consecutive time slots with long-range dependencies. The information not required can be suppressed to improve the accuracy and efficiency of active device detection. 

After passing through the $L$ BiLSTM layers and the attention-based mechanism, the FC layer at the output produces $K$ values corresponding to the total number of devices. Thereby, the output vector $\mathbf{z}^{\text{out}}$ is produced as
\begin{equation}
    \mathbf{z}^{\text{out}} = \mathbf{W}^{\text{out}}   \sum_{l=1}^{L}  \breve{\mathbf{z}}^{[l]}   + \mathbf{b}^{\text{out}},
\end{equation}
where $\mathbf{W}^{\text{out}}$ is the corresponding weight and $\mathbf{b}^{\text{out}}$ the bias, respectively. The softmax layer then maps $K$ output values into $K$ probabilities $( \hat{p}_1, \cdots,  \hat{p}_K  )$ representing the likelihood of being the true support element in the estimated active device support set $\hat{\mathbf{\Upsilon}}$. The $k^{\text{th}}$ probability $\hat{p}_i$ calculated through softmax is given as
\begin{equation} \label{eq:soffy}
    \hat{p}_k = \frac     {  e^{z_k^{\text{out}}}     }      {   \sum_{k=1}^{K} e^{z_k^{\text{out}}}   }.
\end{equation}
Finally, an estimate of the active device support set $\hat{\mathbf{\Upsilon}}$ is obtained by picking from the $K$ elements those having a probability greater than the threshold $\tau$, given as
\begin{equation} \label{eq:eppi}
    \hat{\mathbf{\Upsilon}} = 
\begin{cases}
    1 & \quad \hat{p}_k \geq \tau \\
    0 & \quad \text{otherwise}
\end{cases}
.
\end{equation}
Once $\hat{\mathbf{\Upsilon}}$ in (\ref{eq:eppi}) is obtained, the estimated support $\hat{S}$ is then extracted through the cardinality of the estimated active device support set, i.e., $\hat{S} = | | \hat{\mathbf{\Upsilon}} | |_0$ for the $j$-th time slot. We later show how the estimated active device support set and estimated support are used to evaluate the MUD performance and device identification accuracy.

\section{Model Training, User Detection, and Complexity Analysis}
In this section, we discuss the model training, which in turn is used for signal reconstruction, and find the computational complexity of the proposed attention-based BiLSTM network.

\subsection{Model Training}
During the offline training phase, the network's parameters set $\Theta^{*}$ are computed by minimising the loss function $\mathcal{J}(\Theta)$ (i.e., $\Theta^{*} = \text{arg min}_\Theta \ \mathcal{J}(\Theta)$). During every training iteration, the network parameters are updated using the gradient descent method when the loss function $\mathcal{J}(\Theta)$ is differentiable. Specifically, using the Adam optimiser, the network parameters $\Theta_i$ are updated in the direction of the steepest descent in the $i$-th training iteration, given as
\begin{equation}
    \Theta_{i} = \Theta_{i+1} - \frac{\psi m_{i}}{\sqrt{v_{i} + \epsilon}},
\end{equation}
where $\psi$ is the learning rate determining the step size, and $\epsilon$ is a smoothing term that prevents division by zero. Furthermore, $m_{i}$ and $v_{i}$ are estimates of the mean and uncentered variance of the gradients, respectively, defined as \cite{khan2021transfer} 
\begin{equation}
	\begin{split}
	&m_{i} = \delta_{1} m_{i-1} + (1-\delta_{1}) \nabla 	\mathcal{J}(\Theta_{i})\\[3pt]
	&v_{i}  =\delta_{2} v_{i-1} + (1-\delta_{2}) \nabla 	[\mathcal{J}(\Theta_{i})]^2,
	\end{split}
\end{equation}
where $\delta_{1}$ and $\delta_{2}$ are the decay rates of the moving average. The moving average parameters help in controlling the step size of the optimiser in order to identify the global optimum solution of the training set correctly and prevents the network from looping in a local solution when the training data is not sparse \cite{khan2021transfer}.  

Recalling that the final output of the attention-based BiLSTM network is the $K$-dimensional vector $\hat{\mathbf{p}}$ whose element represents the probability of being the estimated support element from the estimated active device support set $\hat{\mathbf{\Upsilon}}$. In this regard, $\hat{\mathbf{p}}  = [\hat{p}_1 \cdots \hat{p}_K]$ needs to be compared against the true probability $\mathbf{p}$ in the loss function calculation. We employ the cross entropy loss $\mathcal{J}(\mathbf{p}, \hat{\mathbf{p}}, \Theta)$ for network training, defined as \cite{goodfellow2016deep}
\begin{equation}
    \mathcal{J}(\mathbf{p}, \hat{\mathbf{p}}, \Theta) = - \frac{1}{K}    \left ( \sum_{k=1}^{K} p_k \log \hat{p}_k \right) + \lambda \sum_{k=1}^{K} \Theta_k^2,
\end{equation}
where $p_k$ is the ground truth (actual active device), $\hat{p}_k$ is the estimate (estimated active device) of the attention-based BiLSTM network, and $\lambda$ is the $L_2$ regularisation term which is used for weight decaying and in turn, improves the generalisation performance of the network.

\subsection{Blind Data Detection of Active Devices}
In grant-free transmission, the codebook or the spreading sequence is unknown before the signal is detected. This can significantly increase the detection complexity. However, finding the active devices from the received signal can also recover their adopted spreading sequences since a local copy of the spreading sequences is available at the AP. Thus the detection and decoding computational complexity can be reduced significantly while keeping the practical constraints of grant-free NOMA systems intact.

First, AUD is carried out as in (\ref{eq:eppi}) where the estimated active device support set $\hat{\mathbf{\Upsilon}}$, and the estimated sparsity level $\hat{S}$ is obtained from the attention-based BiLSTM network using the stacked received signal $\tilde{\mathbf{y}}$ at the AP. Next, using this estimated active device support set $\hat{\mathbf{\Upsilon}}$, the stacked received signal $\tilde{\mathbf{y}}$ is transformed into a sparse signal $\acute{\mathbf{y}}$, which contains received data for the estimated active devices. Having knowledge of the estimated active devices and their received bits, the spreading sequences employed by these active devices are obtained by selecting the estimated $\hat{S}$ spreading sequences having the highest correlation probability with the spreading sequences at the AP\footnote{Due to the spreading sequences being randomly selected from a pool, there is a possibility of two or more active devices selecting the same spreading sequence, causing spreading sequence collision \cite{kim2021deep}. However, since we have employed complex spreading sequences, and the use of channels of different devices are different, such spreading sequence collision does not have a large impact on the performance of blind detection \cite{3GPP2016DMRS}.} \cite{yuan2017blind}.

\begin{algorithm}[t]
		\caption{The Proposed Attention-based BiLSTM Network.}
		\textbf{Input} Received signal $\hat{\mathbf{y}}$
		
		\textbf{Output} Estimated active user support set $\hat{\mathbf{\Upsilon}}$, estimated sparsity level $\hat{S}$, bits of reconstructed sparse signal $\hat{\boldsymbol{\chi}}$
		
		\textbf{Initialisation}  $\acute{\mathbf{y}} = { 1, \cdots, K  }$
	\begin{algorithmic}[1]

			\Statex{\textbf{Active device support and sparsity estimation}}
			\For{$j = 1$ to $J$} \textbf{do}
			
            \State Obtain $\hat{\mathbf{p}}$ by passing $\hat{\mathbf{y}}^{[j]}$ into the attention-based BiLSTM network
            
            \State $\hat{\mathbf{\Upsilon}}^{[j]} = { k \; | \; \hat{p}_k \geq \tau, 1 \leq k \leq K }$

            \State $\hat{S}^{[j]}$ = $\left| \left| \hat{\mathbf{\Upsilon}}^{[j]} \right| \right|_0$           
            
            \Statex{\textbf{Sparse signal reconstruction}}
            \For{$k = 1$ to $K$} \textbf{do}
            \If {$\hat{\mathbf{p}}_k \geq \tau$}
            \State $\acute{\mathbf{y}}^{[j]}_k = \hat{\mathbf{y}} ^{[j]}_k$
            \Else 
            \State $\acute{\mathbf{y}}^{[j]}_k = 0$
            \EndIf
            \EndFor
            
            \Statex{\textbf{Blind data detection of active devices}}            
            \State $\textbf{w}^T = (  \mathbf{\hat{g}}^T \mathbf{\hat{g}} + \sigma^2 I    )^{-1} \mathbf{\hat{g}}^T$
            
            \State $\hat{\boldsymbol{\chi}}^{[j]} = { \textbf{w}^T \frac{ \acute{\mathbf{y}}^{[j]}}{ \mathbf{\xi}} \text{ mapped to the nearest symbol} }$

            \EndFor

\end{algorithmic}
\textbf{Return} $\hat{\mathbf{\Upsilon}}$, $\hat{S}$, $\hat{\boldsymbol{\chi}}$
\end{algorithm}

Once the spreading sequences employed by the active devices are calculated, blind detection can be carried out. In blind detection, the active device channels are unknown, while the spreading sequences are known. Therefore, based on the sparse signal $\acute{\mathbf{y}}$, which includes the statistical information of channels of all active devices, the blind MMSE weight $\textbf{w}$ can be obtained without the knowledge of device channels. Thereby, the MMSE weight can be calculated as \cite{ameur2020power}
\begin{equation}
    \textbf{w}^T = (  \mathbf{\hat{g}}^T \mathbf{\hat{g}} + \sigma^2 I    )^{-1} \mathbf{\hat{g}}^T,
\end{equation}
where $\mathbf{\hat{g}}$ is the estimated channel between the AP and the devices. After rearranging (\ref{eqReceiveSig}), the transmitted bits of the reconstructed sparse signal $\hat{\mathbf{\boldsymbol{\chi}}}$ for the active devices can then be estimated as
\begin{equation}
    \hat{\boldsymbol{\chi}} = \textbf{w}^T \frac{ \acute{\mathbf{y}}}{ \mathbf{\xi} }.
\end{equation}

By doing so, the active devices' bits are estimated, and the sparse signal $\acute{\mathbf{y}}$ is reconstructed without the need for explicit channel estimation. The entire process is summarised in Algorithm 1.

\begin{table*}[t]
\centering
\caption{Computational complexity comparison for different sparsity levels (the total number of potential devices $K = 200$, the number of subcarriers $N = 100$, the number of hidden layers $L = 3$, width of hidden layer $\alpha = 1000$).}
\label{tabComplexityAna}
\resizebox{\textwidth}{!}{%
\begin{tabular}{|c|c|cccc|}
\hline
\multirow{2}{*}{Technique} & \multirow{2}{*}{Floating point operations (flops)}                                                                                                                 & \multicolumn{4}{c|}{Complexity for different sparsity levels}                                                                                       \\ \cline{3-6} 
                           &                                                                                                                                                                    & \multicolumn{1}{c|}{$S = 10$}           & \multicolumn{1}{c|}{$S = 20$}            & \multicolumn{1}{c|}{$S = 30$}            & $S = 40$            \\ \hline
\textbf{LS-OMP}            & $2S N^2 K + \frac{S^4 + 6S^3 + 7S^2 + 2S}{12} N^3 + S(S+1)N^2 - S$                                                                                                 & \multicolumn{1}{c|}{$1.41 \times 10^9$} & \multicolumn{1}{c|}{$1.76 \times 10^{10}$} & \multicolumn{1}{c|}{$8.15 \times 10^{10}$} & $2.46 \times 10^{11}$ \\ \hline
\textbf{D-AUD}             & \begin{tabular}[c]{@{}c@{}}$2L\alpha^2 + ( 4N + 7L + 2N + 4 ) \alpha + ( S + 3 ) K - \frac{S(S+1)}{2} - 1$\\ $+ 2N + S (\frac{14}{3} N^3 + N^2 - N )$\end{tabular} & \multicolumn{1}{c|}{$5.33 \times 10^7$} & \multicolumn{1}{c|}{$1.00 \times 10^8$}  & \multicolumn{1}{c|}{$1.46 \times 10^8$}  & $1.93 \times 10^8$  \\ \hline
\textbf{LSTM-CS}           & \begin{tabular}[c]{@{}c@{}}$2\alpha^2(4L+3) + 2\alpha(8N+K) +\alpha(3L-1) +3K - 1$\\ $+2N+ S (\frac{14}{3} N^3 + N^2 - N )$\end{tabular}                            & \multicolumn{1}{c|}{$7.87 \times 10^7$} & \multicolumn{1}{c|}{$1.25 \times 10^8$}  & \multicolumn{1}{c|}{$1.72 \times 10^8$}  & $2.19 \times 10^8$  \\ \hline
\textbf{Proposed}          & \begin{tabular}[c]{@{}c@{}}$2\alpha^2(8L+3) + 2\alpha(16N+K) -\alpha(L+3) + 3K-1$\\ $+2N+ S (\frac{14}{3} N^3 + N^2 - N )$\end{tabular}                             & \multicolumn{1}{c|}{$1.04 \times 10^8$} & \multicolumn{1}{c|}{$1.51 \times 10^8$}  & \multicolumn{1}{c|}{$1.97 \times 10^8$}  & $2.46 \times 10^8$  \\ \hline
\end{tabular}%
}
\end{table*}

\subsection{Computational Complexity}
In this subsection, we evaluate the computational complexity of the proposed attention-based BiLSTM network. We evaluate the complexity using the floating-point operations per second (flops) \cite{kim2020deep}, taking into account the complexity of the hidden and deep learning layers of the proposed BiLSTM network at the $j$-th time slot. 

In the first layer of the attention-based BiLSTM network, the input vector has a dimension of $\hat{\mathbf{y}} \in \mathbb{R}^{2N \times 1}$, whereas the weight and bias have the dimensions $\mathbf{W}^{\text{in}} \in \mathbb{R}^{\alpha \times 2N}$ and $\mathbf{b}^{\text{in}} \in \mathbb{R}^{\alpha \times 1}$ respectively. Furthermore, we know that BiLSTM has four gates, which do a forward pass and a backward pass, thereby bringing the generic flop computation per BiLSTM block to
\begin{equation} \label{eqCin}
    \mathcal{C}_{\text{in}} = 8 \times  ( 4N - 1)\alpha + \alpha = 32N\alpha - 7\alpha.
\end{equation}

Next, in the BatchNorm, the element-wise scalar multiplication and addition are carried out twice for normalisation. Thereby, the complexity $\mathcal{C}_{\text{bn}}$ of BatchNorm is given as
\begin{equation}
    \mathcal{C}_{\text{bn}} = 4\alpha.
\end{equation}

Subsequently, in the proceeding hidden layers' BiLSTMs, the hidden weight $\mathbf{W} \in \mathbb{R}^{\alpha \times \alpha}$ is multiplied with the input vector and then the bias term $\mathbf{b} \in \mathbb{R}^{\alpha \times 1}$ is added to it. Next, after passing through the subsequent BatchNorm for each element, the weights are passed through the ReLU activation function. Next, for generalisation, the dropout vector $\mathbf{d}$ is multiplied by the ReLU output $\hat{\mathbf{z}}$. Consequently, the complexity of the $L$ hidden layers $\mathcal{C}_{\text{hide}}$ is given as
\ifCLASSOPTIONonecolumn
\begin{equation}
    \mathcal{C}_{\text{hide}}  = L (  8 \times (2\alpha -1 )\alpha + \alpha + 4\alpha + \alpha +  \alpha)  \;  = 16 L \alpha^2 - L \alpha.
\end{equation}
\else
\begin{equation}
    \mathcal{C}_{\text{hide}}  = L (  8 \times (2\alpha -1 )\alpha + \alpha + 4\alpha + \alpha +  \alpha)  \;  = 16 L \alpha^2 - L \alpha.
\end{equation}
\fi

Following this, the Attention layer performs weighted matrix multiplications with the input and previously sampled data, adds a bias term to the latent data, and multiplies the learnable parameters matrix to compute the scores. Thus, the complexity $\mathcal{C}_{\text{atten}}$ of the Attention layer is
\begin{equation}
    \mathcal{C}_{\text{atten}} = \alpha(2\alpha-1) + 4\alpha^2 + \alpha = 6\alpha^2
\end{equation}

Next, the FC layer at its output has its weights $\mathbf{W}^{\text{out}} \in \mathbb{R}^{K \times \alpha}$ and bias term $\mathbf{b} \in \mathbb{R}^{K\times 1}$ multiplied with the weights from the $L$ hidden layers and the Attention mechanism. Thereby, the FC layer at the output has a complexity $\mathcal{C}_{\text{out}}$ given as
\begin{equation}
    \mathcal{C}_{\text{out}} = (2\alpha - 1)K + K = 2K\alpha.
\end{equation}

The softmax layer computes the $K$ probabilities of potential devices, as in (\ref{eq:soffy}). By doing so, the softmax complexity $\mathcal{C}_{\text{sm}}$ is given as
\begin{equation} \label{eqCend}
    \mathcal{C}_{\text{sm}} = 3K-1.
\end{equation}

From (\ref{eqCin}) to (\ref{eqCend}), the final complexity of the proposed attention-based BiLSTM network is
\ifCLASSOPTIONonecolumn
\begin{equation}
    \mathcal{C}_{\text{ABA}}    = \mathcal{C}_{\text{in}} + \mathcal{C}_{\text{bn}} + \mathcal{C}_{\text{hide}} + \mathcal{C}_{\text{atten}} + \mathcal{C}_{\text{out}} + \mathcal{C}_{\text{sm}}
                         = 2\alpha^2(8L+3) + 2\alpha(16N+K) -\alpha(L+3) + 3K-1.
\end{equation}
\else
\begin{equation}
\begin{split}
    \mathcal{C}_{\text{ABA}}    = & \mathcal{C}_{\text{in}} + \mathcal{C}_{\text{bn}} + \mathcal{C}_{\text{hide}} + \mathcal{C}_{\text{atten}} + \mathcal{C}_{\text{out}} + \mathcal{C}_{\text{sm}}\\
                         = & 2\alpha^2(8L+3) + 2\alpha(16N+K) -\alpha(L+3) + 3K-1.    
\end{split}
\end{equation}
\fi

\indent For an unbiased analysis, we compare the proposed attention-based BiLSTM network with two deep learning-based techniques, D-AUD \cite{kim2020deep} and LSTM-CS \cite{palangi2016distributed}, and a conventional technique, least squares orthogonal matching pursuit (LS-OMP) \cite{shim2012multiuser} for complexity comparison in Table \ref{tabComplexityAna}. In addition, for a fair comparison, the MMSE estimation term has been added to D-AUD and LSTM-CS techniques for signal detection purposes, such that $\mathcal{C}_{MMSE} = 2N + S (\frac{14}{3} N^3 + N^2 - N )$. We examine the computational complexity in flops for different sparsity levels. We observe that the complexity of the proposed attention-based BiLSTM network is slighter higher than D-AUD and LSTM-CS but much lower than that of conventional approaches. This is because the D-AUD technique utilises vanilla FC layers for its network, which do not exploit the temporal correlation of data. Due to this reason, the performance of such networks might degrade with a higher number of active devices. The LSTM-CS uses unidirectional LSTM and therefore has lower computational complexity than the proposed attention-based BiLSTM network. However, as shown in Section V, this results in performance degradation. It is important to note that the complexity of ML-based techniques depends heavily on the network parameters ($L$ and $\alpha$), but not the system parameters (the number of active devices $S$, and the total number of devices $K$). Thus, when $S$ increases from $10$ to $20$, the computational complexity of ML networks increases marginally, but that of LS-OMP increases sharply. Therefore, in a practical grant-free NOMA setting with a higher number of active devices, the ML schemes are more competitive in computational complexity than conventional schemes.

\subsection{Convergence}
\begin{figure}[t]
\centering
\includegraphics[scale=0.5]{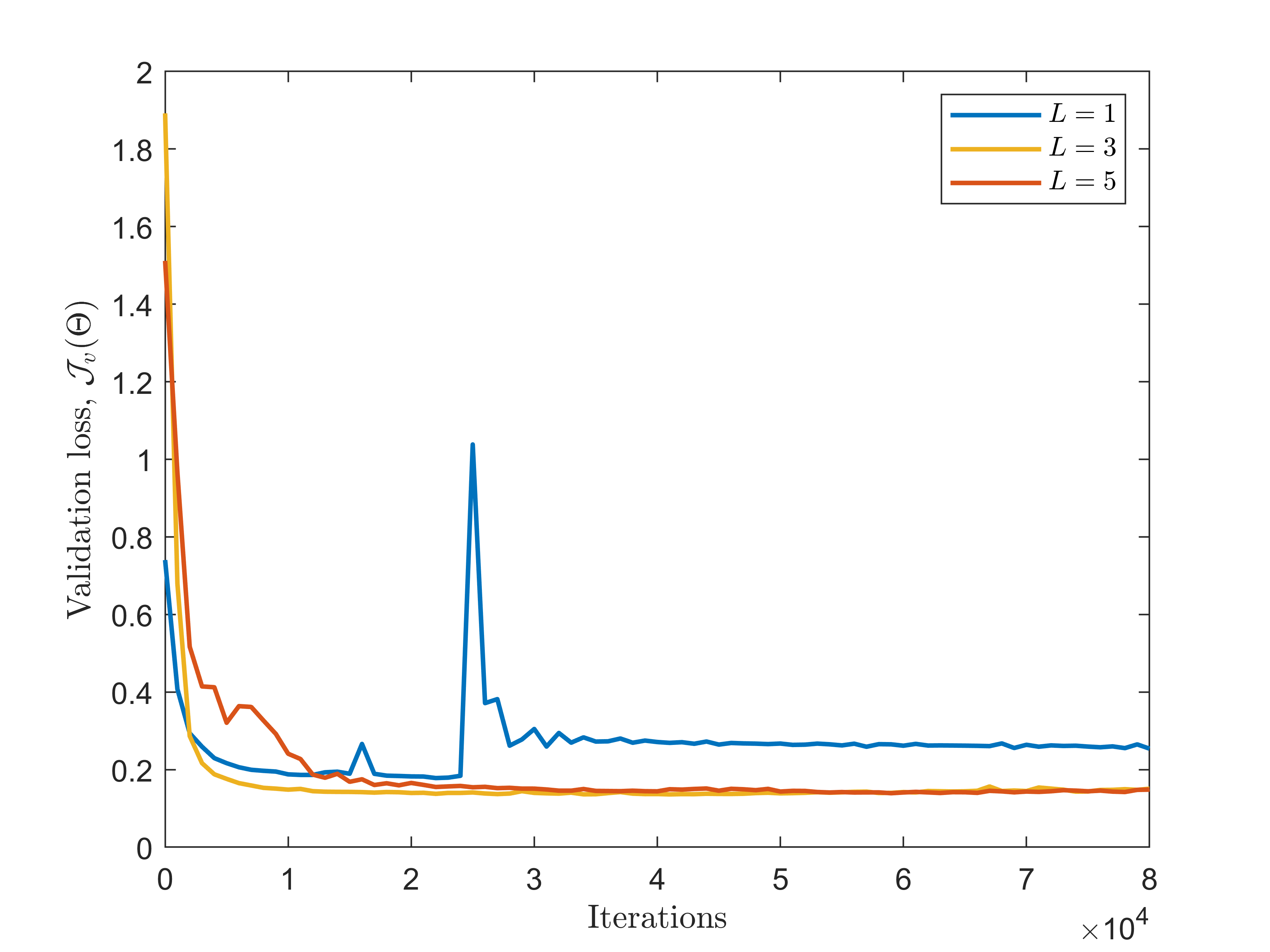}
\caption{Validation loss $\mathcal{J}_v(\Theta)$ for different number of hidden layers $L$, with total number of devices $K = 200$, number of subcarriers $N = 100$, and number of active devices $S = 20$.}
\label{figvalid}
\end{figure}

We examine the validation loss $\mathcal{J}_v(\Theta)$ for a different number of hidden layers $L$ for the proposed network, as shown in Fig.~\ref{figvalid}. We can see that a lower $L$ results in a network being unstable during training, whereas a higher $L$ results in a more stable network but with a slower convergence rate. Thus, we adopt $L=3$ for training dataset generation and also the simulations in this work. Note that the sudden increase in validation loss for the $L = 1$ curve is due to the model overfitting to the training data, causing it to fit noise and outliers and perform poorly on the validation set.

\subsection{Training Dataset Generation} \label{SecDataSet}
In order to determine the optimal network mapping function $g^{*}$ for the stacked received signal $\tilde{\mathbf{y}}$ and support of the $\mathbf{x}$, a comprehensive training dataset is required. A good option in this regard is acquiring a dataset produced using real received signals; however, there is no open-source dataset for the grant-free NOMA scenario at this stage. 

In this work, the training set is generated artificially by sampling values from (\ref{eqReceiveSig}) while keeping the realistic system constraints intact. In essence, during the offline training stage, $U$ training data copies of the stacked received signal vector $\tilde{\mathbf{y}}$ and the support of $\mathbf{x}$ are used as the dataset for network training. With sufficient training data copies, a balanced dataset is obtained, which captures the practical transmission nature of IoT devices. 

The parameter values used for training dataset generation are summarised in Table 4. In particular, in the training dataset, the number of active devices is $S = 20$ among the $K = 200$ potential devices, and the number of subcarriers is $N = 100$. The temporal transmission nature of devices is captured based on (5), and $\eta$ is set to  $0.5$. The length of the time frame is $J = 7$. The number of hidden BiLSTM layers is set as $L = 3$, each with a width of $\alpha = 1000$, each followed by a ReLU activation function. The attention mechanism is placed before the final hidden layer. The output layer is preceded by an FC layer whose width corresponds to the number of classes. The dropout probability  for the dropout layer is set to $\rho_{\text{drop}} = 0.3$. The batch size is set as $20$, while Adam is the optimiser. The value for the latent learning rate $\psi$ is set to $0.001$.

\begin{table}[t]
\centering
\caption{Parameter values used in generating the training dataset.}
\label{tabSimParam}
\resizebox{0.9\linewidth}{!}{%
\begin{tabular}{|l|l|}
\hline
Parameter                                             & Value                                   \\ \hline
Modulation                                            & QPSK                                    \\ \hline
Total devices, $K$                                    & $200$                                   \\ \hline
Total subcarriers, $N$                                & $100$                                   \\ \hline
Active devices, $S$                                   & $20$                                 \\ \hline
Channel, $g_{k}$                                      & Rayleigh fading, $\mathcal{CN}(0, 1)$   \\ \hline
Multiple access signature                             & Random selection                        \\ \hline
Time slots, $J$                                       & $7$                                     \\ \hline
Temporal correlation, $\eta$                          & $0.5$                                   \\ \hline
SNR distribution                                      & $0$ dB to $20$ dB                       \\ \hline
Detection threshold, $\tau$                           & $0.5$                                   \\ \hline
Hidden layers, $L$                                    & $3$                                     \\ \hline
Hidden layer width, $\alpha$                          & $1000$                                  \\ \hline
Activation layer, $\sigma$                            & ReLU                                    \\ \hline
Learning rate, $\psi$                                 & $0.001$                                 \\ \hline
Batch size, $B$                                       & $20$                                    \\ \hline
Dropout probability, $\rho_{drop}$                    & 0.3                                     \\ \hline
Validation split                                      & $20\%$                                  \\ \hline
Moving average decay rate, $\delta_{1}$, $\delta_{2}$ & $\delta_{1} = 0.9$, $\delta_{2} = 0.99$ \\ \hline
\end{tabular}%
}
\end{table}

\section{Results and Discussion}
In this section, we evaluate the performance of the proposed attention-based BiLSTM network in solving the MUD problem. We also plot the performance of four benchmark solutions: two traditional CS solutions, LS-OMP \cite{shim2012multiuser} and dynamic CS-based MUD method \cite{wang2016dynamic},  one ML-based LSTM-CS MUD method \cite{palangi2016distributed} and the Oracle least squares (LS) algorithm.

The motivation for considering these four benchmarks is as follows. We consider the LS-OMP as it is the standard CS technique that is always considered as one of the benchmarks in this research field. The dynamic CS-based and ML-based LSTM-CS methods are considered because they take temporal correlation into account during MUD. Additionally, the ML-based LSTM- CS method demonstrates the advantage gained from considering the proposed BiLSTM over vanilla LSTM. The Oracle LS algorithm is considered as it provides the theoretical performance lower bound, although it is impractical in real-world situations where perfect knowledge is unavailable.

For a fair comparison, we make the following assumptions in the implementations of the four benchmark schemes:
\begin{itemize}
    \item For the two traditional benchmark solutions, the sparsity level is assumed to be known at the AP due to the assumption of the channels being perfectly known; only the sparse support location is unknown at the AP.
    
    \item For the ML-based LSTM-CS MUD method, the core working of the method is adopted from \cite{palangi2016distributed}, but the LSTM layer is adapted to our architecture (as in Fig. (2)) for a fair comparison.
    
    \item For the ML-based LSTM-CS MUD method, we assume that it does not need any channel state information, i.e., it is unaware of the sparsity level, sparse support location and the channels.
    
    \item For the Oracle LS algorithm, we assume perfect knowledge of the channel state information, user sparsity level and sparse support location.

\end{itemize}

In the simulations, unless otherwise stated, $K = 200$ potential devices simultaneously share $N = 100$ orthogonal resource blocks. Thus, the overloading factor\footnote{The overloading factor is defined as the ratio of the number of potential devices to the number of available resource blocks in the system, i.e., overloading factor (\%) = $\frac{K}{N} \times 100$.} is 200\%. The number of active devices is in the range $S= 10-40$. We employ $M = 4$-ary complex spreading sequences, where both the real and imaginary parts take values from the set $\{ -2, -1, 0, 1\}$. For every time slot, there are $S$ number of active devices, where the active device support set $\mathbf{\Gamma}^{[j]}$ in each time slot has $S/2$ devices transmitting in the next time slot, i.e., $\eta = 0.5$, while the remaining are randomly selected from $\{ 1, 2, \cdots, K \}$. The number of time slots is fixed at $J = 7$ to conform to the LTE-Advanced protocol \cite{ETUR2021evolved}. The signals being transmitted are modulated by Quadrature Phase Shift Keying (QPSK). Furthermore, all channels are assumed to follow an independent Rayleigh fading, and the channel fading coefficient is generated following $g_{n,k} \sim  \mathcal{CN}(0, 1)$. The path loss between the AP and the $k$-th device is modeled as $128.1 + 37.6 \log_{10} (d_i)$, where $d_i$ is the distance (in km) \cite{ETUR2021evolved}. The results are averaged over $1000$ Monte Carlo trials.

\subsection{Performance Metrics}
In order to appropriately evaluate the performance, including the quality of support estimation, device identification, and multi-device data detection, we use the following metrics: the detection probability ($\rho_d$), the accuracy, and the average bit-error rate (BER) as performance metrics. Given the bits of the reconstructed sparse signal $\hat{\boldsymbol{\chi}}^{[j]}_k$ for device $k$ at the $j$-th time slot, the performance metrics are defined as follows.
\begin{itemize}

    \item Detection probability: This metric evaluates the performance of support estimation. It is defined as the ratio of the number of detected active devices to the number of \textit{all} active devices, given as
    \begin{equation}
        \rho_d = \frac{1}{S} \displaystyle\sum_{k \in \Gamma^{[j]}} \mathds{1}_{ { \hat{\boldsymbol{\chi}}}^{[j]}_k \neq 0 } .
    \end{equation}

    \item Accuracy: This metric evaluates the performance of the quality of support estimation for device identification. It is defined as the ratio of the number of correctly identified active devices to the number of \textit{all} active devices, expressed as a $\%$ and given as
    \begin{equation}
        \text{Accuracy (\%)} = \frac{1}{S} \displaystyle\sum_{k \in \Gamma^{[j]}} \mathds{1}_{ { \mathbf{\Gamma}^{[j]} == \hat{\mathbf{\Upsilon}}^{[j]} } } \times 100.
    \end{equation}
    
    \item Average BER: This metric evaluates the performance of multi-device data detection. It is defined as the ratio of incorrectly recovered bits transmitted by the active devices to all bits transmitted by the active devices. It should be noted that the average BER includes a penalty for decoding the wrongly detected active devices.
    
\end{itemize}

\begin{figure*}[t]
\begin{subfigure}{0.45\textwidth}
\centering
\includegraphics[width=\textwidth]{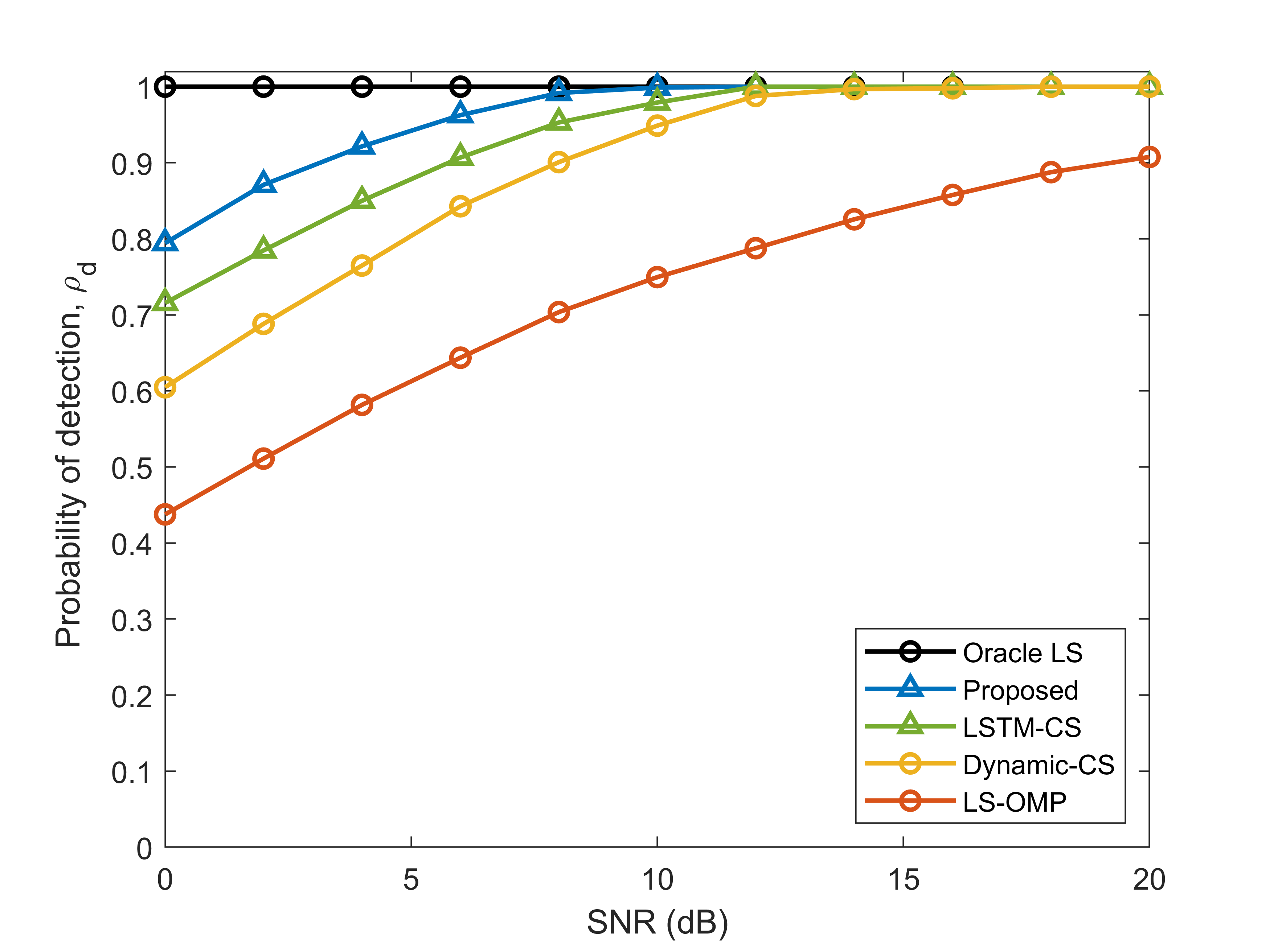}
\caption{$S = 10$}
\end{subfigure}
\quad
\begin{subfigure}{0.45\textwidth}
\centering
\includegraphics[width=\textwidth]{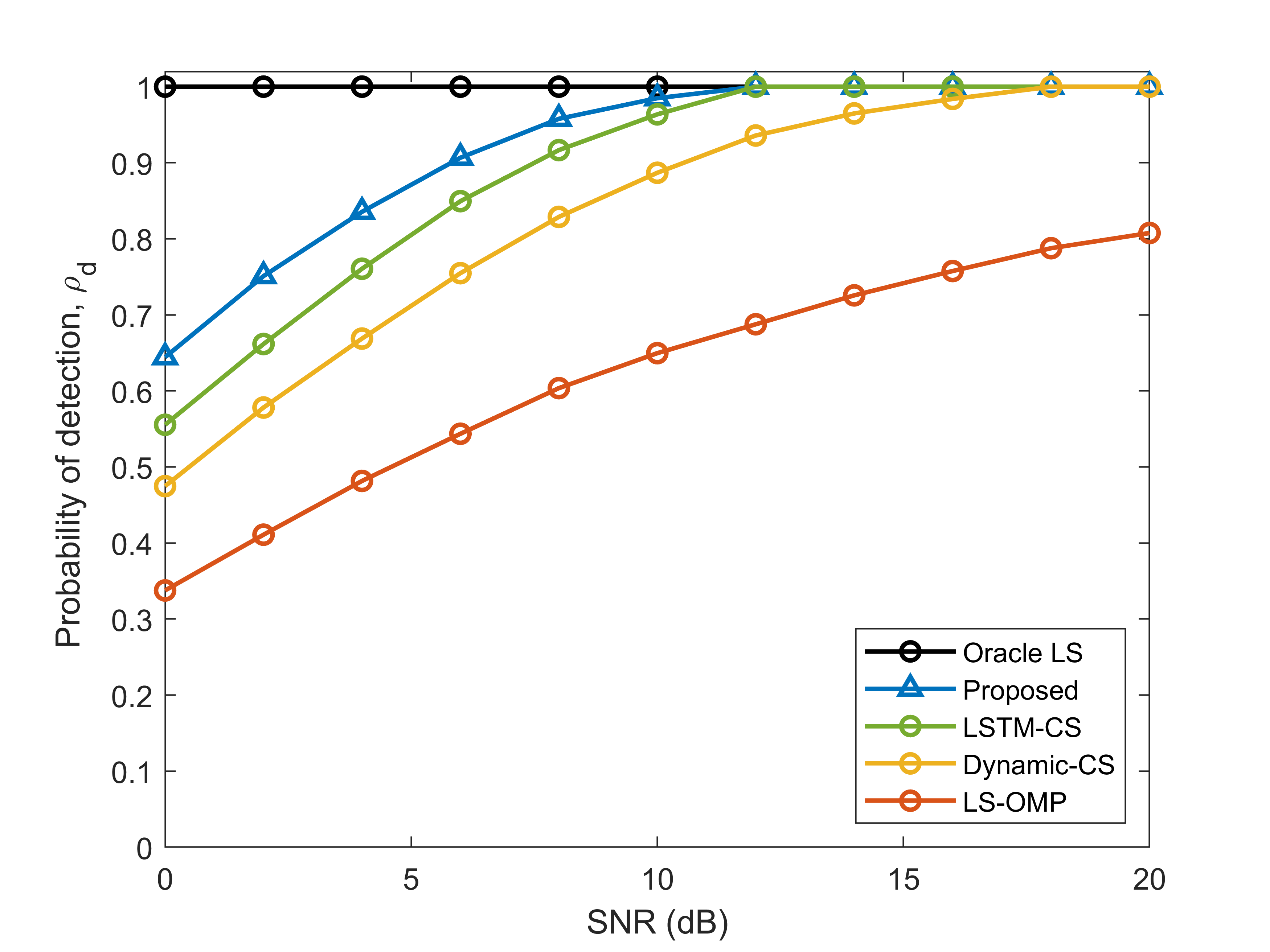}
\caption{$S = 20$}
\end{subfigure}
\caption{Probability of detection, $\rho_d$, versus SNR (dB) for the number of active devices $S$, with the total number of potential devices $K = 200$, and the number of subcarriers $N = 100$.}
\label{figProbDectS}
\end{figure*}

\subsection{Support Estimation}
Fig.~\ref{figProbDectS} plots the detection performance, $\rho_d$, versus the SNR (dB) for $S = 10$ and $S = 20$, with $K = 200$, and $N = 100$. The following trends can be observed from the figure. The Oracle LS gives the theoretical best performance (100\% detection probability for the considered scenario), which is the same for all SNR values. As the SNR increases, the performance of all the schemes slowly approaches to that of the Oracle LS. The LS-OMP performs the worst since it ignores the temporal correlation in the device activation history. The dynamic CS-based MUD method performs better than LS-OMP since it considers the temporal correlation in the device activation history. The ML-based LSTM-CS method performs better than the two traditional algorithms but cannot perform similarly to the proposed BiLSTM network due to its unidirectional architecture. The proposed attention-based BiLSTM network outperforms all these benchmark algorithms, i.e., it exhibits a higher detection probability of successfully identifying the correct number of active devices against all other schemes. For instance, the proposed attention-based BiLSTM network achieves the Oracle LS detection performance at SNR $= 8$ dB and SNR $= 12$ dB, respectively, for $S = 10$ and $S = 20$ active devices. It should be noted that the proposed attention-based BiLSTM network is unaware of the device sparsity level and detects the active devices based on the received signal only, compared to other traditional algorithms, which are based on the assumption of the known channels and device sparsity level. As the number of active devices $S$ increases from $10$ to $20$, the detection performance of the proposed attention-based BiLSTM network decreases gradually. The decrease in performance is attributed to the introduction of additional interference, variability, and overlapping patterns. These complexities pose challenges for the model to effectively capture and learn the underlying patterns and relationships within the data\footnote{Note that in order to further enhance the network detection performance, data augmentation techniques can be introduced to control the variations and to improve the model robustness, enabling it to capture complex relationships better. This is outside the scope of this work.}.

\begin{table}
\centering
\caption{Device identification accuracy versus the number of active devices $S$, with the total number of potential devices $K = 200$, the number of subcarriers $N = 100$, and SNR $= 6$ dB.}
\label{tabDevIdent}
\begin{tabular}{|c|c|c|c|c|} 
\hline
\multirow{2}{*}{\begin{tabular}[c]{@{}c@{}}\# of Active\\ Devices\end{tabular}} & \multicolumn{4}{c|}{Accuracy (\%)}        \\ 
\cline{2-5}
                                                                                & LS-OMP & Dynamic-CS & LSTM-CS & Proposed  \\ 
\hline
10                                                                              & 71.92  & 84.57      & 90.25   & 94.85     \\ 
\hline
15                                                                              & 64.10  & 79.29      & 86.51   & 92.54     \\ 
\hline
20                                                                              & 59.75  & 77.47      & 80.09   & 84.84     \\ 
\hline
25                                                                              & 44.40  & 72.74      & 73.84   & 74.99     \\ 
\hline
30                                                                              & 36.52  & 47.94      & 54.16   & 62.38     \\ 
\hline
35                                                                              & 4.29   & 32.84      & 40.21   & 46.97     \\ 
\hline
40                                                                              & 0      & 15.14      & 22.52   & 29.31     \\
\hline
\end{tabular}
\end{table}

\subsection{Device Identification}
Device identification can help the AP prioritise service provision considering the available resources and provide access to devices based on their priority. Table \ref{tabDevIdent} shows the accuracy of correctly identified active devices at $K = 200$, $N = 100$, and SNR $= 6$ dB. It should be noted again that the traditional schemes in this regard assume complete knowledge of the device sparsity level and that their accuracy is based on identifying the actual active device support set only. On the contrary, the proposed attention-based BiLSTM network follows a practical approach where the active device sparsity level is first estimated. Then, the actual active device support set is identified based on the estimated sparsity level. 

We can see from the figure that the trends between the various benchmark schemes are the same as in Fig. 5. The proposed attention-based BiLSTM network outperforms the benchmark schemes by correctly identifying the actual active device support set with higher accuracy. The ML-based LSTM-CS method cannot correctly identify all the active devices because it relies on forward direction architecture only. On the contrary, due to its forward and reverse direction architecture, the proposed BiLSTM network can identify more active devices correctly. It can be seen that with the increasing number of active devices, the accuracy of correctly identifying the actual active device support set decreases, which is to be expected when grant-free NOMA systems operate in overloaded conditions.

\begin{figure}[t]
\centering
\includegraphics[scale=0.5]{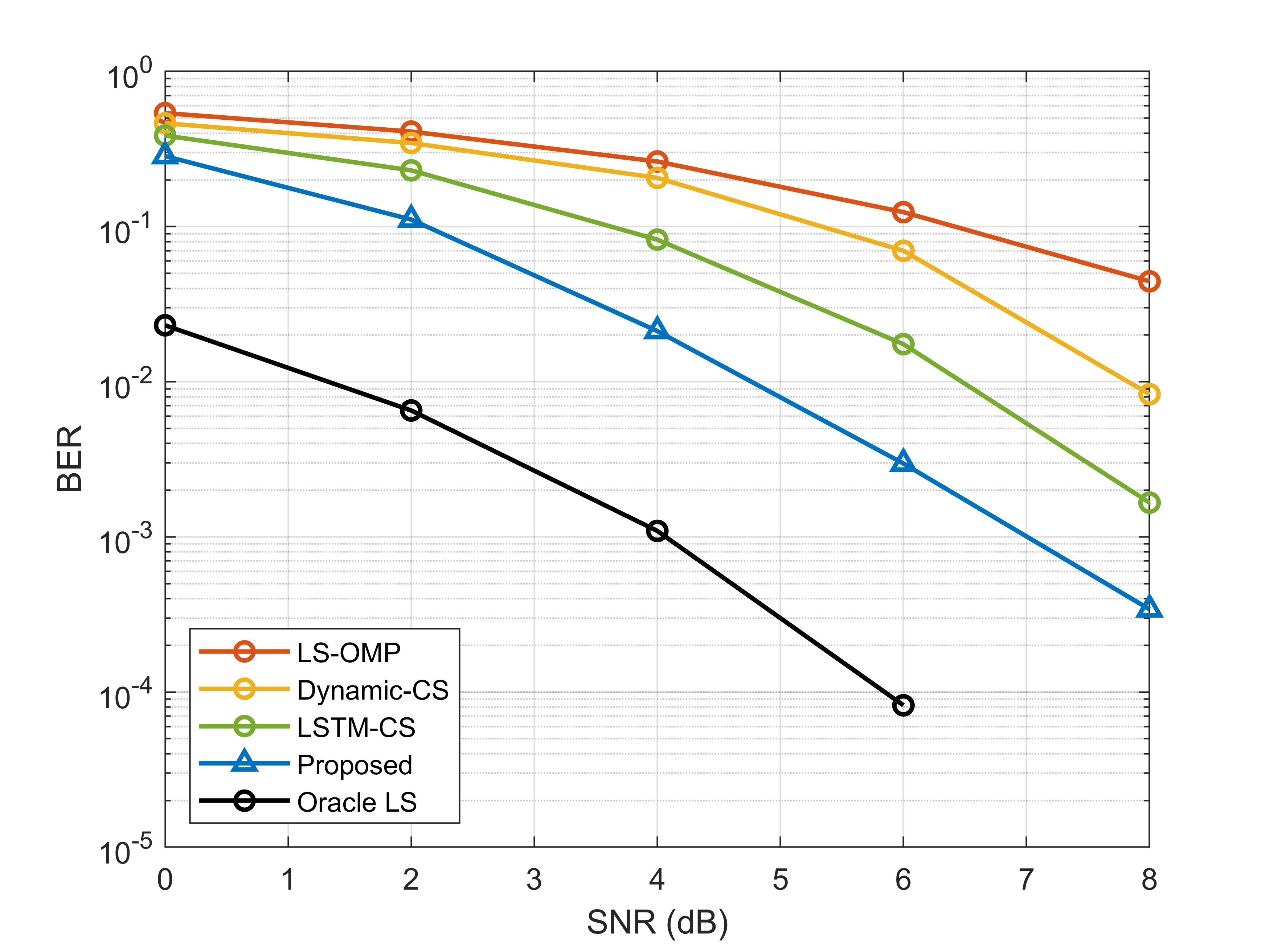}
\caption{Average BER versus the SNR (dB), with the total number of potential devices $K = 200$, the number of subcarriers $N = 100$, and the number of active devices $S = 20$.}
\label{figBerversusSnr}
\end{figure}

\subsection{Multi-User Data Detection}
Fig.~\ref{figBerversusSnr} plots the average BER of the considered algorithms against the SNR (dB), with $K = 200$, $N = 100$, and $S = 20$. In all scenarios, our proposed attention-based BiLSTM network outperforms the benchmark schemes over the whole considered range of SNR, including the ML-based LTSM-CS method. For SNR $> 4$ dB, the gap between the proposed attention-based BiLSTM network and the Oracle LS algorithm is about $3$ dB only. This performance gap with the Oracle LS algorithm is because it fully assumes the active device sparsity level and active device support set. The inaccurate active device estimation causes the performance gap as a side effect of the grant-free NOMA system.

\begin{figure}[t]
\centering
\includegraphics[scale=0.5]{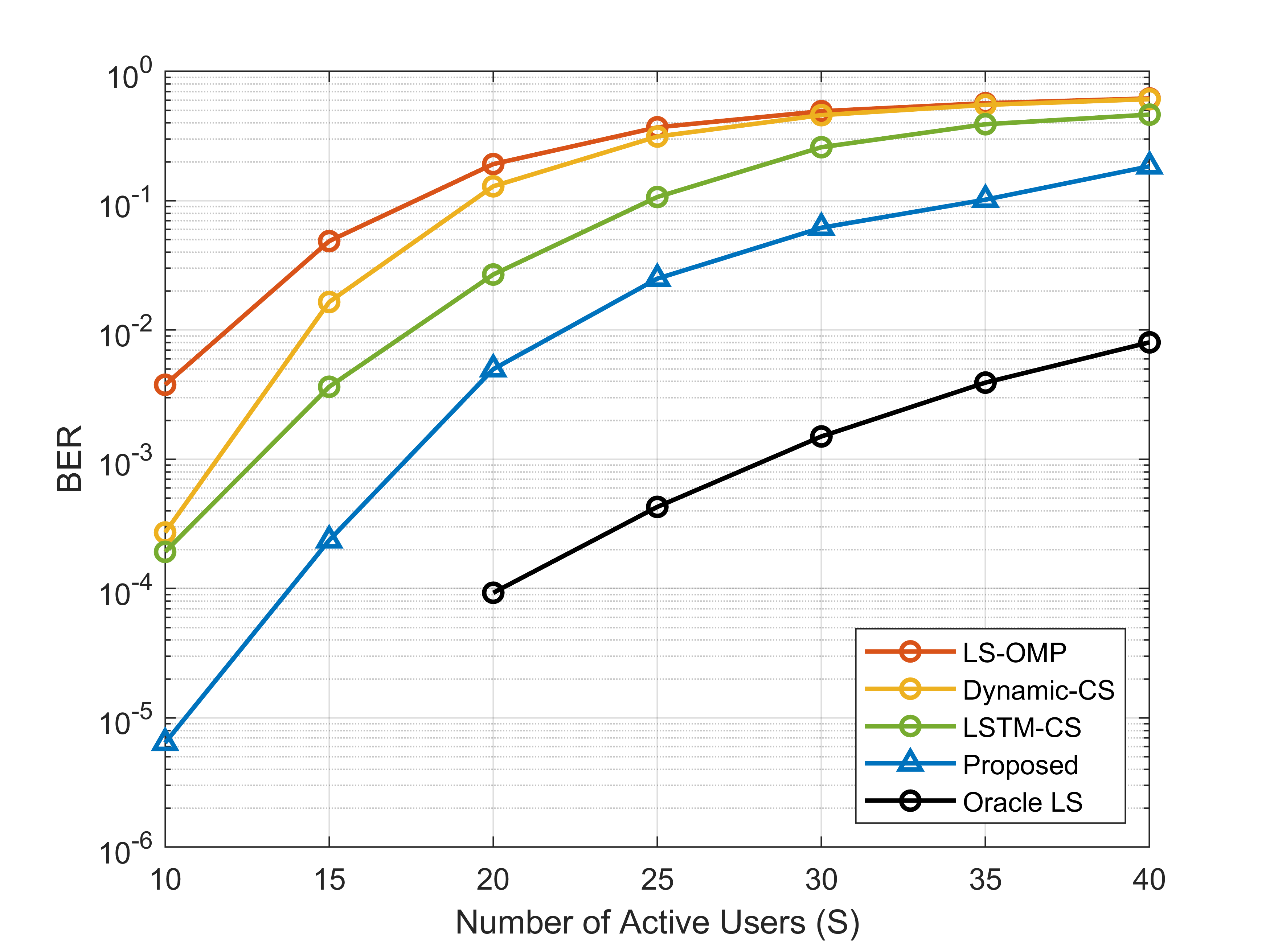}
\caption{Average BER versus the number of active devices $S$, with total number of potential devices $K = 200$, the number of subcarriers $N = 100$, and SNR $= 6$ dB.}
\label{figBerversusActiveUsers}
\end{figure}

Fig.~\ref{figBerversusActiveUsers} plots the average BER against the active device sparsity $S$, with $K = 200$, $N = 100$, and SNR = $6$ dB. Unlike the computational complexity of the proposed network in Section IV-C, the BER performance is impacted by the number of active users. For all methods, the BER decays as the active devices increase. Even so, the proposed attention-based BiLSTM network exhibits consistently lower BER than the benchmark schemes throughout the whole considered range of SNR. The ML-based LSTM-CS method performs better than traditional methods initially but saturates with a high number of active devices since it cannot capture their temporal activation pattern due to its unidirectional architecture. The consistent performance gains of the proposed attention-based BiLSTM network show that the network has precisely mapped the underlying relationship between device activity and received signals, given that the network is trained for $S = 20$ active devices.

\begin{figure}[t]
\centering
\includegraphics[scale=0.5]{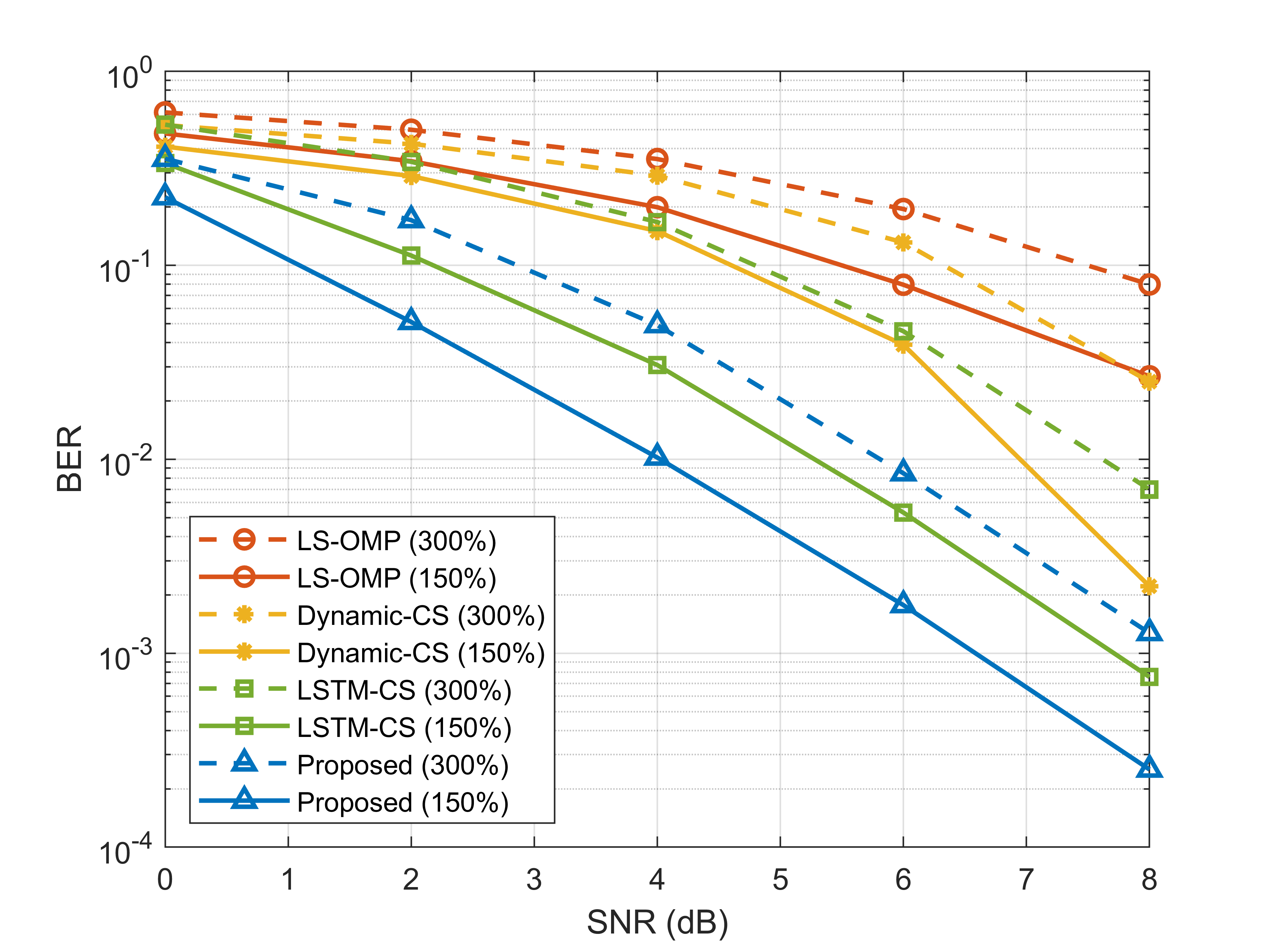}
\caption{Average BER versus SNR (dB) for varying overloading factor, with number of subcarriers $N = 100$, and number of active devices $S = 20$.}
\label{figBerversusTotalUsers}
\end{figure}

Fig.~\ref{figBerversusTotalUsers} plots the average BER against the SNR (dB) for varying overloading factors, with $N = 100$ and $S = 20$. It is evident that the average BER for all benchmark techniques increases with a higher overloading factor as the potential devices $K$ are increased, making the system prone to correlation errors. Even so, the average BER of the proposed attention-based BiLSTM network compared to conventional techniques is lower, manifesting that the proposed attention-based BiLSTM network can load more devices with the same training configuration. This is because the proposed attention-based BiLSTM network has higher tolerance and robustness against increased overloading factors due to decoupled correlated activation patterns.

\begin{figure}[t]
\centering
\includegraphics[scale=0.5]{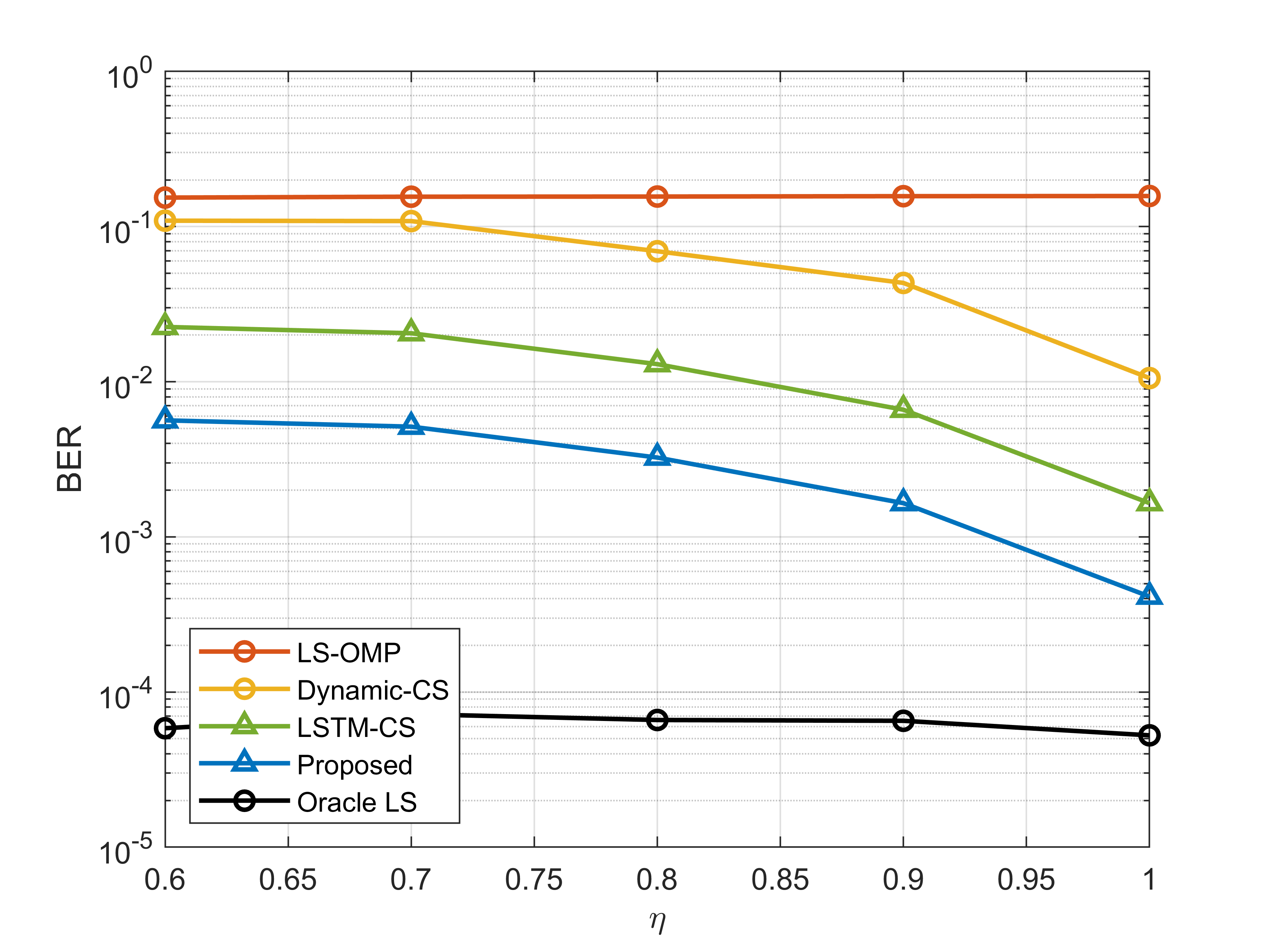}
\caption{Average BER versus the temporal correlation parameter $\eta$, with total number of devices $K = 200$, number of subcarriers $N = 100$, number of active devices $S = 20$, and SNR $= 6$ dB.}
\label{figBerversusTemporalCorr}
\end{figure}

Fig.~\ref{figBerversusTemporalCorr} plots the average BER against the temporal correlation parameter, $\eta$, with $K = 200$, $N = 100$, $S = 20$ and SNR = $6$ dB. Note that the result for $\eta = 1$ corresponds to the special case of frame-wise joint sparsity, i.e., devices' activity remains constant over an entire data frame. We can see that the proposed network performs well for all values of $\eta$. Herein, the LS-OMP algorithm performs poorly because it does not utilise the extra information present in the previous time slots for temporal activity. Conversely, the BER of the dynamic CS-based method is also relatively higher due to its dependence on devices' activity in the $(j-1)$ time slot only. The ML-based LSTM-CS method performs better than the dynamic CS-based method because it takes the temporal activity of devices in all time slots. However, because the ML-based LSTM-CS utilises a forward direction LSTM only, it does not completely capture the activation pattern of active devices. On the contrary, it can be seen that the increasing temporal correlation parameter $\eta$ enhances the BER performance of the proposed attention-based BiLSTM network. This is because the proposed attention-based BiLSTM network has bidirectional LSTM units, which successfully capture the underlying mapping of the stacked received signal $\tilde{\mathbf{y}}$ with the temporal correlation of device activity between different time-slots using the estimated support of $\mathbf{x}$. This further testifies to the generability of the proposed attention-based BiLSTM network in different transmission patterns. The Oracle LS algorithm outperforms the proposed algorithm and remains consistent since it assumes a complete active device support set.

\subsection{Discussion on Robustness, Scalability and Generalisation}
The results in Figs. 5-9 show that the proposed attention-based BiLSTM network, which is trained on $S = 20$, $N = 100$, $K = 200$ and $\eta = 0.5$, is robust to changes in the key system parameters. We can see that the trained BiLSTM network still performs well when there is a change in the number of active devices (Figs. 5 and 7), the number of potential devices or, equivalently, the overloading factor (Fig. 8) or temporal correlation model (Fig. 9), and does not need to be retrained for the considered practical range of considered values ($10 \leq S \leq 40$, $0.5 \leq \eta \leq 1$ and $150 \leq K \leq 300$). This is because training the network at $\eta = 0.5$ allows it to learn the important features of the device activation patterns, and it still performs well when the parameters change. This shows that the proposed network is generalisable to different system parameters. 

In addition, the proposed network is a good solution for grant-free NOMA systems to provide faster access for massive IoT devices. As demonstrated in Table~\ref{tabComplexityAna}, the proposed network's computational complexity is comparable to the state-of-the-art ML-based solution and does not heavily depend on the system parameters. Thus, when the number of active devices increases, or the number of potential devices in the system becomes large, the computational complexity increases only marginally. Thus, the proposed scheme is scalable and is suitable for faster access in massive IoT device scenarios.

\section{Conclusions}
In this paper, we proposed an attention-based BiLSTM network for AUD in an uplink grant-free NOMA system by exploiting the temporal correlation of active user support sets. First, a BiLSTM network is used to create a pattern of the device activation history in its hidden layers, whereas the attention mechanism provides essential context to the device activation history pattern. Then, the complex spreading sequences are utilised for blind data detection without explicit channel estimation from the estimated active user support set. Thus, the proposed mechanism is efficient and does not depend on impractical assumptions, such as prior knowledge of active user sparsity or channel conditions. Through simulations, we demonstrated that the proposed mechanism outperforms several existing benchmark MUD algorithms and maintains lower computational complexity. In this work, we have applied the proposed framework to spreading based grant-free NOMA scheme. Future work can investigate the generalisation of the proposed framework to other signature-based grant-free NOMA schemes.

\bibliographystyle{IEEEtran}
\bibliography{IEEEabrv, references}

\end{document}